\documentclass[fleqn,usenatbib]{mnras}

\usepackage[T1]{fontenc}
\usepackage{ae,aecompl}
\usepackage{array}
\usepackage{todonotes}
\usepackage{float}

\usepackage{graphicx}	
\usepackage{amsmath}	
\usepackage{amssymb}	

\title[NS-WD mergers]{Faint Rapid Red Transients from \\ Neutron Star - CO White Dwarf Mergers}

\author[Y.Zenati, A.Bobrick \& H.B.Perets]{
Yossef Zenati$^{1}$, 
Alexey Bobrick$^{2}$
and
Hagai B. Perets$^{1}$
\\
$^{1}$Physics Department, Technion - Israel Institute of Technology,
Haifa 3200004, Israel\\
$^{2}$Lund University, Department of Astronomy and Theoretical physics, Box 43, SE 221-00 Lund, Sweden
}

\date{Accepted XXX. Received YYY; in original form ZZZ}
 
\pubyear{2019}

\begin{document}
\label{firstpage}
\pagerange{\pageref{firstpage}--\pageref{lastpage}}
\maketitle

\begin{abstract}
Mergers of neutron stars (NS) and white dwarfs (WD) may give rise to observable explosive transient events. We use 3D hydrodynamical (SPH) simulations, as well as 2D hydrodynamical-thermonuclear simulations (using the FLASH AMR code) to model the disruption of CO-WDs by NSs, which produce faint transient events. We post-process the simulations using a large nuclear network and make use of the SuperNu radiation-transfer code to predict the observational signatures and detailed properties of these transients. We calculate the light-curves (LC) and spectra for five models of NS - CO-WD  mergers. The small yields of $^{56}{\rm Ni}$ (few$\times10^{-3}\,{\rm M_{\odot}}$) result in faint, rapidly-evolving reddened transients (RRTs) with B (R) - peak magnitudes of at most $\sim-12$ ($-13$) to $\sim-13$ ($-15$), much shorter and fainter than both regular and faint/peculiar type-Ia SNe. These transients are likely to be accompanied by several months-long, $1$--$2$ mag dimmer red/IR afterglows. We show that the spectra of RRTs share some similarities with rapidly-evolving transients such as SN2010x, although RRTs are significantly fainter, especially in the I/R bands, and show far stronger Si lines. We estimate that the upcoming Large Synoptic Survey Telescope could detect RRTs at a rate of  up to $\sim10-70$ yr$^{-1}$, through observations in the R/I bands. The qualitative agreement between the SPH and FLASH approaches supports the earlier hydrodynamical studies of these systems.
\end{abstract}




\section{Introduction}

\begin{table*}
\begin{centering}
\begin{tabular}{|c|c|c|c|c|c|c|c|c|c|c|}
\hline 
\# & ${M_{\rm WD}}[M_\odot]$ & ${\rho_{\rm max}[{\rm g}/{\rm cm}^3]}$  & ${R_{0}/r_{t}}$ & ${\rm \%He_{4}}$ & ${\rm \%C_{12}}$  & ${\rm \%O_{16}}$ & ${E_{\rm K} [{\rm erg}]}$ & ${\rm IGE_{U}[{\rm 10^{-3}M_{\odot}}]}$ & ${M_{\rm Tot-U}[{\rm 10^{-3}M_{\odot}}]}$\tabularnewline
\hline 
C & ${0.55}$ & ${8.5\times10^{6}}$ & ${0.8}$ & ${-}$ & ${50}$ & ${50}$& ${2.8\times10^{48}}$ & ${9.0}$ & ${19}$\tabularnewline
\hline 
D & ${0.63}$ & ${6.4\times10^{6}}$ & ${1}$ & ${9}$ & ${50}$ & ${41}$& ${3.6\times10^{49}}$ & ${9.6}$ & ${210}$\tabularnewline
\hline 
E & $0.63$ & $8.5\times10^{6}$ & $0.8$  & $4$ & $49$ & $47$ & $1.4\times10^{49}$ & $6.1$ & $90$\tabularnewline
\hline 
F {*} & $0.63$ & $2.3\times10^{6}$ & $1.1$ & $-$ & $50$ & $50$ & $1.5\times10^{48}$ & $3.3$ & $8.1$\tabularnewline
\hline 
J & $0.8$ & $4.4\times10^{7}$ & $0.8$ & $-$ & $50$ & $50$ & $3.2\times10^{49}$ & $45$ & $94$\tabularnewline
\hline 
\end{tabular}
\par\end{centering}
\caption{\label{tab:WD-models} The initial parameters of the simulated NS-WD merger
models. The columns show the names of the models, the masses of the WDs ($M_{\rm WD}$) in solar masses, the maximum density in the torus in our FLASH simulations ($\rho_{\rm max}$), the torus radius in units of the tidal radius ($R_0/r_t$) and the initial mass fractions of ${\rm He}_4$, ${\rm C}_{12}$ and ${\rm O}_{16}$. The last three columns show the kinetic energy at the end of the FLASH simulations and the total mass of iron-group ejecta (IGE) and all of the ejecta ($M_{\rm tot-U}$). The energy powering the outflows comes almost entirely from accretion and not the nuclear processes involved, e.g. \citet{Zen+19a}. \protect \\
{*}These models were run with ${M_{\rm NS}=2M_{\odot}}$\protect \\
}
\end{table*}

\begin{figure*}
\includegraphics[width=0.36\linewidth]{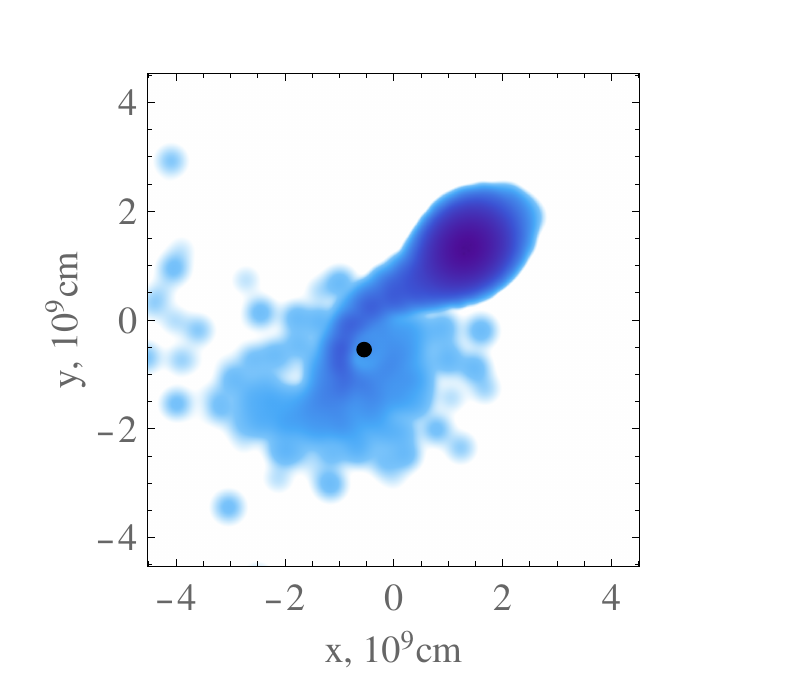}
\includegraphics[width=0.36\linewidth]{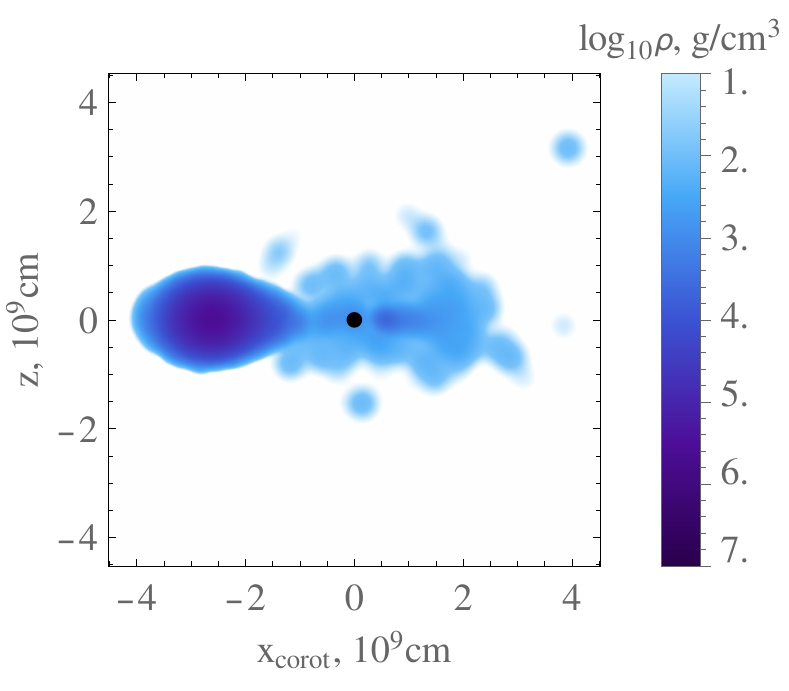}

\includegraphics[width=0.36\linewidth]{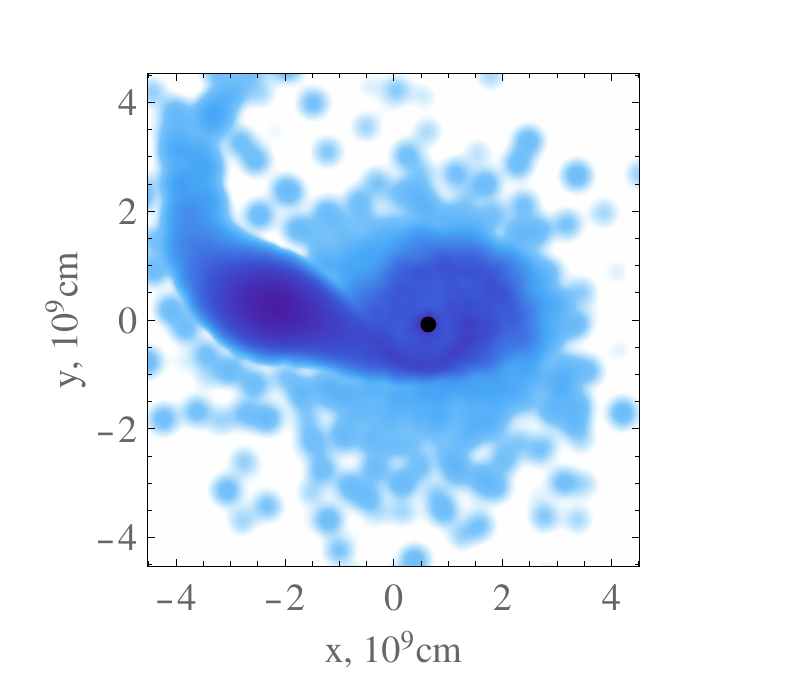}
\includegraphics[width=0.36\linewidth]{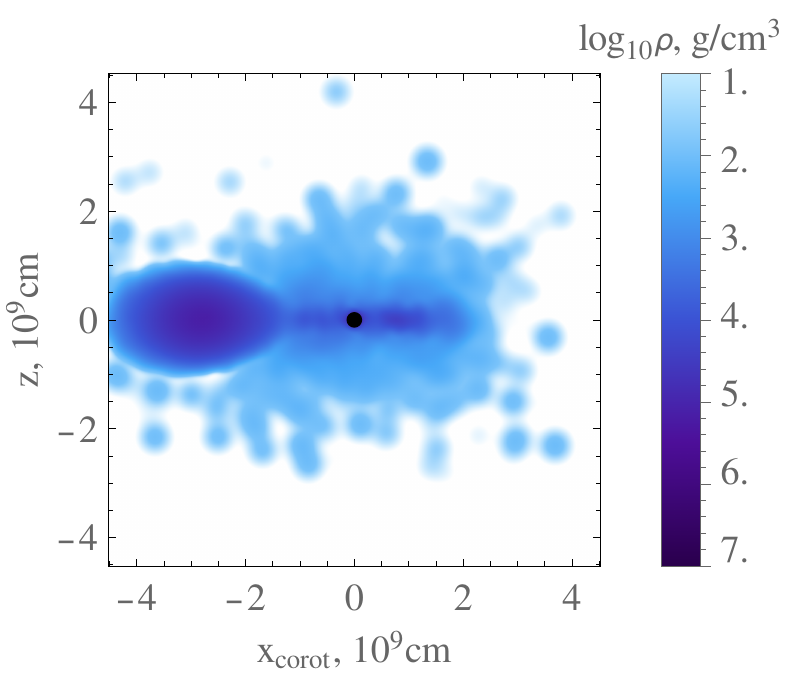}

\includegraphics[width=0.36\linewidth]{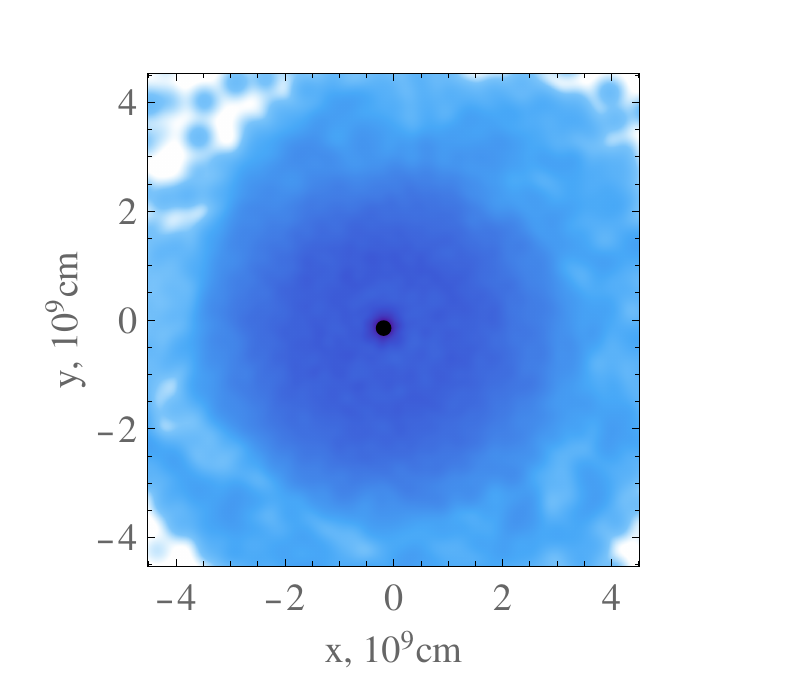}
\includegraphics[width=0.36\linewidth]{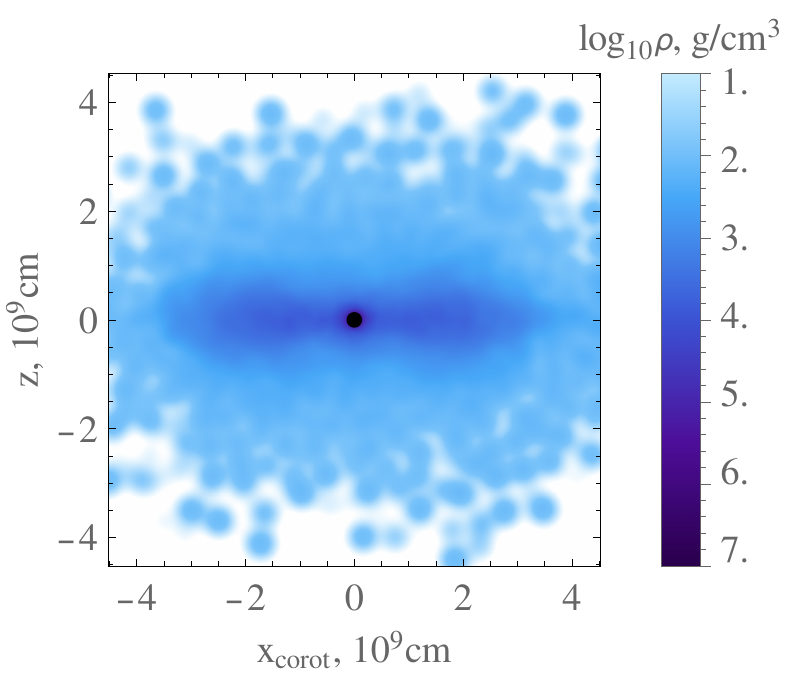}
\caption{Snapshots for density distribution from the SPH simulation of model D. The left column shows the density in the horizontal $x$-$y$ plane, and the right column shows the density distribution in the vertical plane $x_{\rm corot}$-$z$, corotating with the bulk of the white dwarf mass and centered on the NS. The snapshots are taken at $34.4\,\textrm{s}$, $111.7\,\textrm{s}$ and $303.6\,\textrm{s}$ from the beginning of the simulation. The simulation starts with the donor relaxed into the state when the mass-transfer stream has already developed. As the mass-transfer rate grows over time, the donor gets increasingly deformed by the tidal field. At a certain point (here, at $100\,{\rm s}$) the donor stops being gravitationally bound, gets stretched by the tidal field and quickly relaxes into a debris-disk state (here, by about $300\,{\rm s}$). The shock heating caused by self-crossings of the material during this stage is responsible for most nuclear burning in the simulation.}
\label{fig:SPH}
\end{figure*}

\begin{figure*}
\includegraphics[width=0.45\linewidth]{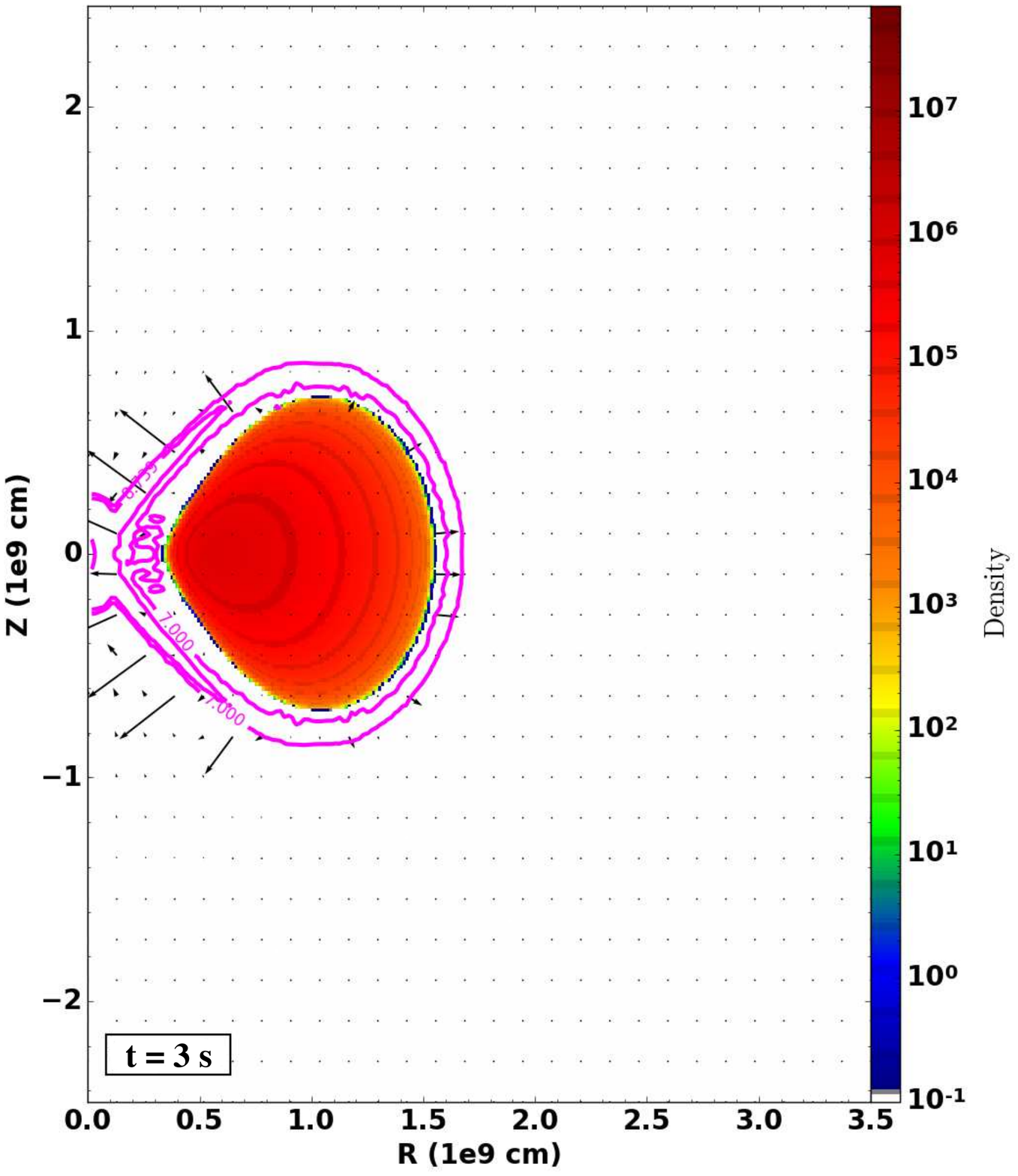}
\includegraphics[width=0.45\linewidth]{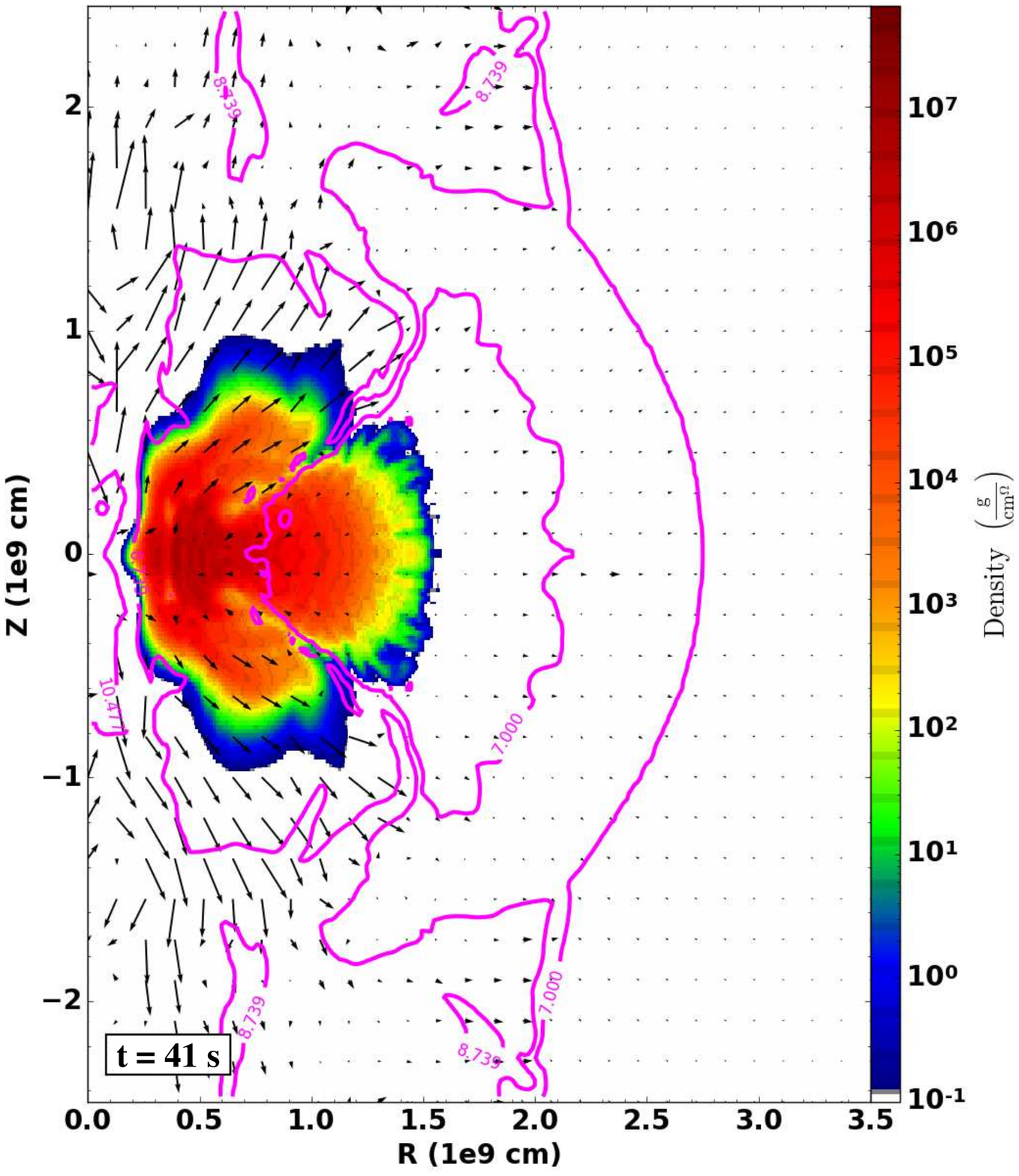}
\includegraphics[width=0.45\linewidth]{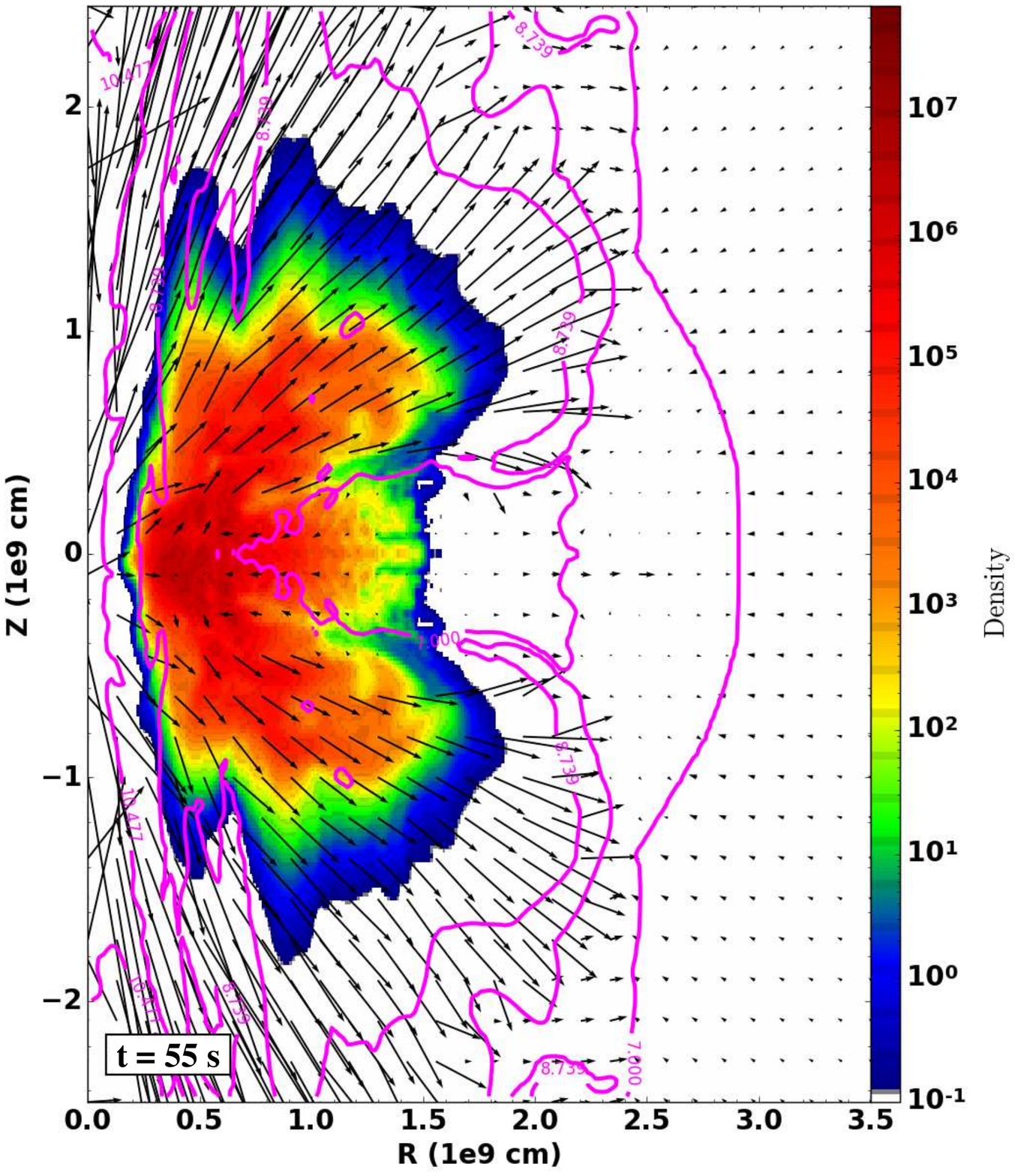}
\includegraphics[width=0.45\linewidth]{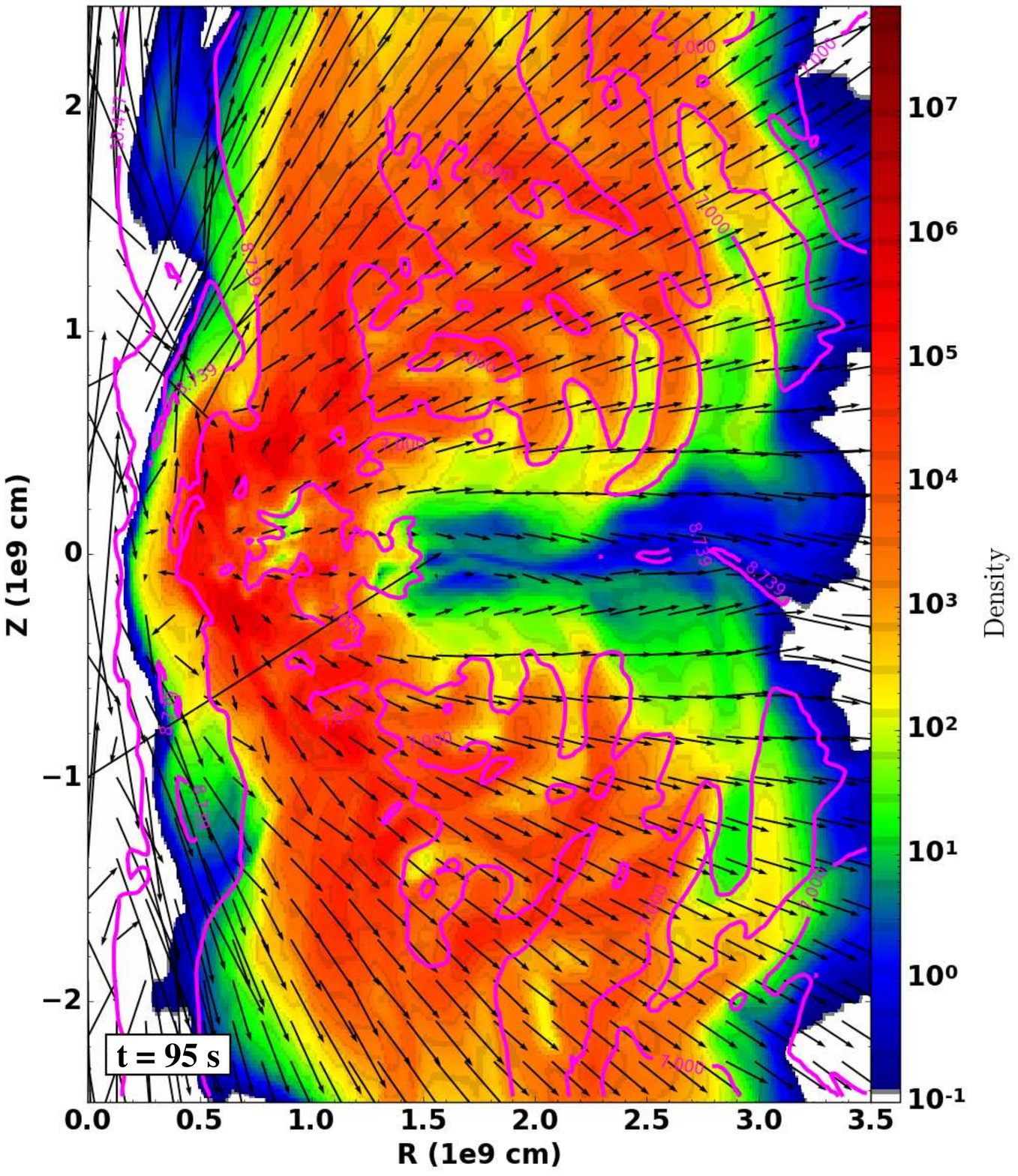}
\caption{The evolution of the WD debris for model E based on FLASH simulations. The panels show the colour coded density distribution and velocity fields (black arrows) throughout the simulation. The velocity scale is $4500\,{\rm km}\cdot{\rm sec}^{-1}$. Magenta lines show the surfaces of constant temperature.
}
\label{fig:FLASH}
\end{figure*}


Mergers of neutron stars (NSs) or black holes (BH) with white dwarfs (WDs) have been explored through several hydrodynamical and nuclear-hydrodynamical simulations, e.g. \citet{FM13, Bobrick2017,Zen+19a,FMM19}. All CO WD-NS mergers are expected to lead to unstable mass transfer \citep{Bobrick2017} where the WD is tidally disrupted on dynamical timescales and forms an extended debris disk around the NS \citep{10.1093/mnras/208.4.721,1999ApJ...520..650F}. The disc evolves mostly viscously but is also affected by nuclear burning. Disc viscosity and nuclear burning together power outflows throughout the disc \citep{2012MNRAS.419..827M}. Nuclear burning in the disc proceeds steadily (with a possible weak detonation) and produces small amounts of  $^{56}{\rm Ni}$ (at most $10^{-2}$ ${\rm M_{\odot}})$, e.g. \citet{Zen+19a}. Such a small amount of $^{56}{\rm Ni}$ is expected to lead to optical transients which are much fainter than ordinary type Ia supernovae (SNe; which produce more than an order of magnitude more  $^{56}{\rm Ni}$). In this paper, we present the first synthetic LCs and spectra for CO WD-NS mergers and show that they may give rise to a novel class of faint, rapid red transients (RRTs), with no evidence for hydrogen or helium.


To model the merger, we use two approaches. The first approach makes use of 2D coupled hydrodynamical-thermonuclear simulations with the FLASH code, similar to the method introduced by \citet{2013ApJ...763..108F}. In these simulations, the NS-WD merger is followed only from the advanced stage at which the debris disk of the disrupted WD has already formed and given rise to an axisymmetric structure which can be explored in 2D \citep[][hereafter Paper I]{Zen+19a}. In the second approach, we applied a 3D SPH code to follow the full NS-WD merger, including the early pre-disruption phase, but with nuclear burning treated only in a post-processing step. As we discuss below, these two approaches are complementary and generally produce very similar results.


As we and others have already found \citep{2013ApJ...763..108F,Zen+19a}, the evolution of the merger and the evolution of the WD debris disk are dominated by the viscous evolution and the gravitational energy, with nuclear burning playing only a minor role in the structural evolution of the disk and the outflows, but playing an important role in driving the luminosity evolution of the transient. These results motivated us to use 3D SPH simulations which did not include any feedback from nuclear burning, which were then post-processed using a 125-elements nuclear network to analyze nuclear burning yields in such mergers. The nuclear yields and observational predictions could then be compared with our results from the 2D FLASH simulations.

To make detailed observational predictions for the transients arising from NS-WD mergers, we have mapped the outputs from our hydrodynamic runs, post-processed with a large nuclear network (in the SPH case), onto an input grid for a 1D radiation transport code. We subsequently compared the obtained LCs and spectra to LCs and spectra of various classes of peculiar transients. As we show below, WD-NS mergers produce faint RRTs, fainter and redder than any of the observed SNe thought to arise from thermonuclear explosions, and may, therefore, give rise to a new class of potentially observable transients. These transients may be accompanied by an even fainter red/infrared several months-long afterglow.

Below we describe our numerical simulations and the various types of initial conditions we explored in Section 2. We then present the main results in Section 3 and discuss and summarize them in Section 4.

\section{Methods}

Our original FLASH simulations and their results are described in detail in \citet{Zen+19a} (Paper I). Here we briefly summarize the main ingredients of the model: the WD profiles produced with the MESA stellar evolution code, the simulations of the debris disk evolution with the FLASH code, and the nuclear post-processing - all of which we already described in Paper I. We then focus on the new SPH simulations using the Water code and then describe our radiation-transfer modelling with the SuperNu code and the resulting observational signatures (light-curves and spectra), which are the main focus of the current paper.

\subsection{The neutron star and white dwarf models}
\label{sec:StModels}

We describe the properties of each of the NS-WD models we simulated in Table \ref{tab:WD-models}. We obtained the structural properties of the WDs through detailed stellar evolution models of single and binary stars
using the MESA code \citep{2011ApJS..192....3P,Paxton2013,2015ApJS..220...15P}.
In all cases, we considered only stellar progenitors of solar metallicity. Our models include both typical CO WDs as well as hybrid HeCO WDs. The former are the outcomes of regular evolution of single stars, which results in WDs composed of $\sim$50$\,\%$ carbon and $\sim$50$\,\%$ oxygen. The hybrid WDs, containing both CO and He, are the outcomes of \emph{binary} stellar evolution, as described in \citet{Zen+18}. Subsequently, we used these stellar profiles to set up our WD models in the SPH code and  their bulk compositions to set up the disrupted WD's in the FLASH code.

We do not resolve the NSs in our simulations and model them as point masses. We considered two NS masses, $1.4\,M_{\odot}$ and $2\,M_{\odot}$, which corresponds to the typical range of NS masses observed in binary radio pulsars, e.g. \citet{Lattimer2012}.

\subsection{FLASH simulations}

During the early stages of disruption, the NS tidally shreds the WD into a debris disc, e.g \cite{10.1093/mnras/208.4.721,1999ApJ...520..650F}. The detailed initial conditions and the structure and evolution of the modelled disk can be found in Paper I.

We simulate the evolution of such a disc using the publicly available FLASH code v4.2 \citep{2000ApJS..131..273F}. We employ the unsplit ${\rm PPM}$ solver of FLASH hydrodynamics code in ${\rm 2D}$ axisymmetric cylindrical coordinates on a grid of size ${\rm 1\times1\left[10^{10}cm\right]}$ using adaptive mesh refinement. We follow the approach similar to the one adopted in other works on thermonuclear SNe (e.g. \citealt{Mea+09}). We handle detonations by the reactive hydrodynamics solver in FLASH without the need for a front tracker, which is possible since unresolved Chapman--Jouguet detonations retain the correct jump conditions and propagation speeds. 

In the cases of NS - HeCO-WD mergers (models D and E), we identify the ignition and detonation of helium in the accretion disk.  The helium ignition occurs at sufficiently high accretion rates at densities of  $\rho \gtrsim 8\times10^{5}\,{\rm g}/{\rm cm}^{-3}$, which, through supersonic burning, gives rise to a (weak) He-detonation in the HeCO mixed layer.

In the cases of NS - CO-WD mergers, the required pressure and density for a weak detonation are $P\gtrsim 2\cdot 10^{25} \, {\rm erg} \cdot {\rm cm}^{-3}$ and $\rho \gtrsim 2.5-3 \times10^{7}\, {\rm g}\cdot {\rm cm}^{-3}$, respectively, which is higher than in the case of NS-He CO mergers. The timescale for ${\rm C}_{12}$ burning is also much longer, in effect leading to a different and delayed (by about one second) detonation compared to the HeCO WD cases.

In this work, we have also checked the resolution convergence following \citet{2019ApJ...876...64F} and have not observed any changes in the evolution of the disk. We evolved our simulations for  $(175$-$270)\,{\rm s}$. In one test case, we ran a lower-resolution model for up to $400\,{\rm s}$.  In comparison, the orbital timescale at the circularisation radius, in case of model D, is about $40$ seconds. The viscous timescale at the circularisation radius $R_0$, given by:
\begin{equation}
t_{\rm visc}\sim 260{\rm\,s} \left(\frac{0.1}{\alpha}\right)\left(\frac{R_{0}}{10^{9.3}\rm cm}\right)^{3/2}
\left(\frac{1.4M_{\sun}}{M_{\rm c}}\right)^{1/2}\left(\frac{H_{0}}{0.5R_{0}}\right)^{-2},
\end{equation}
is about $260$ seconds, assuming the alpha-disc model parameter $\alpha=0.1$, as adopted in our FLASH simulations. In the equation, $M_{\rm c}$ is the central mass of the NS and $H_0$ is the scale height in the disc. In the inner regions of the disc, where most nuclear burning takes place, due to a smaller radius, the viscous timescale is only a few tens of seconds, which is well covered by our runs.

While our FLASH simulations do not capture all the details of the initial 3-dimensional onset phase of mass transfer in the binaries, the orbital and viscous timescales of the disc are well captured by our models. The simulation timescales also agree with the timescales of our SPH simulations, which also last for several hundreds of seconds. The overall similarity of the nuclear abundance patterns and a reasonable, typically up to $1\,{\rm dex}$, agreement of the nuclear yields supports the picture that both codes capture the phase which is most relevant to nuclear burning.

As discussed in detail in Paper I, we use the Helmholtz EOS \citep{2000ApJS..126..501T}, and account for self-gravity of the disk. We employ a $19$-isotope reaction network with correct treatment of the burning front \citep{1989BAAS...21.1209F}. To prevent the production of artificial unrealistic early detonation that may arise from insufficient numerical resolution, we applied a limiter approach following \citet{2013ApJ...778L..37K}. Further details of these simulations can be found in Paper I. The initial conditions and the outcomes of the simulations we use in the current paper are summarized in Table \ref{tab:WD-models}.

\subsection{SPH simulations}

We have complemented and verified our modelling with the FLASH code by using a smoothed particle hydrodynamics (SPH) code Water \citep{Bobrick2017}. We set up the SPH simulations and their post-processing aiming to reproduce the same physical mergers like the ones modelled by the FLASH method. The SPH approach allows us to evolve the models from early stages of mass transfer in 3D and follow the disruption of the WD and the formation of the debris disk, while our 2D FLASH simulations can only be initiated at a late stage when a WD debris disk is already assumed to have formed. On the other hand, our SPH models only include nuclear burning as a post-processing step. The two approaches, therefore, are different and complement each other.

The Water code is based on the most up-to-date SPH prescriptions available and derives from the Oil-on-Water code used to model the onset stages of mass transfer \citep{Bobrick2017}, see also \citet{Church2009}. We refer to \citet{Bobrick2017} for full implementation details and summarise the key components of the code below.

We solve the Navier-Stokes equations by discretizing the Lagrangian as done in \citet{Springel2010}. This way, we ensure exact energy and momentum conservation, only limited by the accuracy of the gravity solver, which we implement as in \citet{Benz1990}. We limit the maximum radius of SPH particles in a Lagrangian and continuous way, thus partially mitigating the so-called fall-back problem when ejected particles become large and acquire too many neighbours upon falling back on the donor. We use Wendland W6 kernels \citep{2012MNRAS.425.1068D}, which prevent pairing instability by construction, and set the number of neighbours to $400$. We use the artificial viscosity from \citet{2010MNRAS.408..669C}, which switches on the viscosity only near shocks and does not damp sound waves. We base the artificial conductivity on the prescription from \citet{Bobrick2017, Hu2014}, this way reducing discontinuities in thermal energy near shocks. Finally, we use the KDK integrator as described in \citet{1997astro.ph.10043Q,2005MNRAS.364.1105S,2010MNRAS.408..669C} with a time step limiter by \citet{2009ApJ...697L..99S}, which improves shock treatment.

We constructed the SPH models by using the stellar profiles prepared with the MESA code, described in Section~\ref{sec:StModels} (Table~\ref{tab:WD-models}). We made the SPH models of the WDs by mapping SPH particles onto appropriately spaced spherical shells \citep{saff1997distributing, Bobrick2017,Raskin2016}.  We used SPH models with equal particle masses to avoid numerical artefacts reported in earlier studies, e.g. \citet{LorenAguilar2009,Dan2011}. We relaxed the single models in isolation over six dynamical times of the WD. Then we placed the donors in a binary with the NS at $1.8\,a_{\rm RLOF}$ and relaxed the binary in a co-rotating frame, continually spiralling in the donor down to $0.975\,a_{\rm RLOF}$ over two orbital periods. Then we relaxed the binary at the final separation over an additional half a period, to start the main simulation with an existing mass-transfer stream, see \citet{Bobrick2017}. We removed the small number of particles accreted onto the NS during the relaxation stage. Subsequently, we ran the simulations for up to $10$ orbital periods into the WD disruption until after the formation of a developed disc. Since the SPH simulations allow us to model the merger from the early stages to the advanced stages when a fully developed disc has formed, we expect that they capture most of the phases relevant to nucleosynthesis. Furthermore, as we show in Section~\ref{sec:MergDyn}, most of the nuclear material, according to both FLASH and SPH, is produced in the advanced stages of the merger when the disc has fully formed, which is captured by both codes.

In the SPH simulations, we used the same Helmholtz equation of state as we used in FLASH, but did not include feedback from nuclear burning. However, as mentioned above, the latter likely has only a little effect on the overall dynamical evolution of the disk \citep{Zen+19a}. We softened the gravity of the NS by a W6 Wendland kernel with a core radius of $0.1\,R_{\rm WD}$ and have checked the convergence by simulating models in $20\,{\rm K}$, $50\,{\rm K}$ and $100\,{\rm K}$ resolution.

\subsection{Nuclear post-processing}
We use a 19-isotope $\alpha$-chain network in the FLASH simulations and do not evolve chemical compositions in the SPH simulations. The $\alpha$  network can adequately capture the energy generated during nuclear burning \citep{2000ApJS..126..501T}. To verify the abundances produced by the FLASH code and to calculate the detailed nuclear yields for the SPH runs, we applied a nucleosynthetic post-processing step with a large network. For the grid-based FLASH simulations, we made use of $4000$-$10000$ tracer particles to follow the evolution and to track the composition, velocity, density, and temperature of the WD debris. These particles were evenly spaced throughout the WD-debris disk with a step of ${2\times10^{8}\,{\rm cm}}$. The detailed histories of density and temperature from tracer particles (from FLASH) and SPH particles (from the SPH Water code) were post-processed with MESA (version 10390) one-zone burner \citep{2015ApJS..220...15P}. We employed a 125-isotope network that includes neutrons and composite reactions from JINA REACLIB \citep{2010ApJS..189..240C}. We subsequently used the outputs of the original $\alpha$ network in the FLASH case, to correctly weigh the nuclear abundances over the material, and the outputs of the detailed MESA PPN network in the SPH case for further analysis.

~

~

~

~

\subsection{Radiation transfer modeling using SuperNu}
\label{sec:RTSetup}

Following the nuclear post-processing, we mapped the physical properties from this step as inputs for the openly available radiation transfer code SuperNu \citep{2013ApJS..209...36W,2014ApJS..214...28W} in order to calculate the light-curves (LC) and spectra expected from the mergers. SuperNu uses Implicit Monte Carlo (IMC) and Discrete Diffusion Monte Carlo (DDMC) methods to stochastically solve the special-relativistic radiative transport and diffusion equations to first order in $v/c$ in three dimensions. The hybrid IMC and DDMC scheme used in SuperNu makes it computationally efficient in regions with high optical depth. This approach allows SuperNu to solve for energy diffusion with very few approximations, which is very relevant for supernova light curves.
    
SuperNu code follows the free-expansion phase of supernovae using a velocity grid. We map the 3D velocities of the FLASH material and the SPH particles to a 1D-spherical velocity grid of SuperNu, as they are, at the end of simulations. In other words, we set the masses in each of the grid cells of SuperNu equal to the sum of the masses of the corresponding elements (grid cells for FLASH, particles for SPH). We set the chemical compositions and electron fractions $Y_e$ as the mass-weighted average compositions in each cell. Unlike the Eulerian-based FLASH code, SuperNu can handle true vacuum within cells. Consequently, zero-mass grid cells containing no mapped particles require no special treatment, although the velocity distributions did not contain any significant gaps. 

Since most of the material is gravitationally bound at the end of the simulations, both in FLASH and SPH, mapping the 3D velocities directly to SuperNu represents fast dynamical ejecta and is an approximation, as we discuss in Section~\ref{sec:Disc}. To check for the effect of the additional ejecta released as a slow wind, we additionally initialised SuperNu runs by randomly assigning the element velocities typical values of $400\,{\rm km}/{\rm s}$ with a $1$-$\sigma$ spread of $200\,{\rm km}/{\rm s}$. As we discuss in Section~\ref{sec:Disc} the true fractions of material in fast and slow ejecta are uncertain, and thus we obtain the characteristic signal expected from both types of ejecta.

We initialised the SuperNu code at $0.5\,{\rm d}$ after the merger. The homologously expanding material adiabatically cools down by at least six orders of magnitude in temperature by this point. We set the temperature of the material initialised in SuperNu to upper estimate value of $10^5\,\textrm{K}$. Setting initial temperatures to values lower than $10^5\,\textrm{K}$ led to essentially the same lightcurves or spectra. We have also verified that the input distributions for SuperNu from SPH and FLASH simulations broadly looked similar.

\begin{table*}
\begin{centering}
\begin{tabular}{|c|c|c|c|c|c|c|c|c|c|c|c|}
\hline 
\# & ${M_{\rm WD} [M_{\odot}}]$ & ${^{56}{\rm Ni}_{{\rm B}} \ [\times10^{-3} {M_{\odot}}]}$ & ${\rm B_{peak}}$ & $\rm {\rm \varDelta B_{15}}$ & ${\rm R_{peak}}$ & ${\rm \varDelta R_{15}}$& ${\rm I_{peak}}$ & ${\rm \varDelta I_{15}}$\tabularnewline
\hline 

C & ${0.55}$ & ${5.25} \ (6.51)$ & $ -12.81 \  (-12.68) $ & $4.50 \ (2.85) $ & $-14.70 \ (-14.66)$ & $3.29\ (1.81)$ & $-14.93 \ (-14.89)$ & $2.34 \ (1.78)$\tabularnewline
\hline 
D & ${0.62}$ & ${6.16 \ (10.03)}$ & $ -12.80 \ (-12.66) $ & $3.41 \ (2.85)$ & $-14.72 \ (-15.11)$ & $2.75 \ (1.54)$ & $-14.98 \ (-15.26)$ & $2.01 \ (1.57)$\tabularnewline
\hline 
E & $0.62$ & ${2.84 \ (10.14)}$ & $ -12.02 \ (-12.92)$ & $3.48 \ (1.93)$ & $-14.23 \ (-14.90)$ & $2.52 \ (1.44)$ & $-14.75 \ (-15.01)$ & $2.10 \ (1.40)$\tabularnewline
\hline 
F & $0.62$ & ${3.38 \ (8.82)}$ & $ -12.19 \ (-12.40)$ & $3.64 \ (2.47)$ & $-14.36 \ (-14.79)$ & $2.77 \ (1.51)$ & $-14.80 \ (-14.91)$ & $2.02 \ (1.47)$\tabularnewline
\hline 
J & $0.8$ & ${1.78 \ (26.16)}$ & $ -10.88 \ (-12.28)$ & $3.35 \ (2.67)$ & $-13.21 \ (-14.35)$ & $1.91 \ (1.52)$ & $-13.87 \ (-14.85)$ & $1.82 \ (1.42)$\tabularnewline
\hline 
\end{tabular}
\par\end{centering}
\caption{\label{tab:WD-models2} The outcomes of the merger models explored in this study, based on FLASH simulations and SPH results in close brackets. The columns correspond to: model number (1); the white dwarf mass (2); the amount of $^{56}{\rm Ni}$ produced (3); the B-peak luminosity (4); the decrease in B-magnitude over 15 days since the peak ($\Delta {\rm B}_{15}$) (5); the R-peak luminosity (6); the decrease in R-magnitude over 15 days since the peak (7); the I-peak luminosity (8) and the decrease in I-magnitude over 15 days since the peak (9).
\protect \\
 }
\end{table*}

\begin{figure}
\includegraphics[scale=0.16]{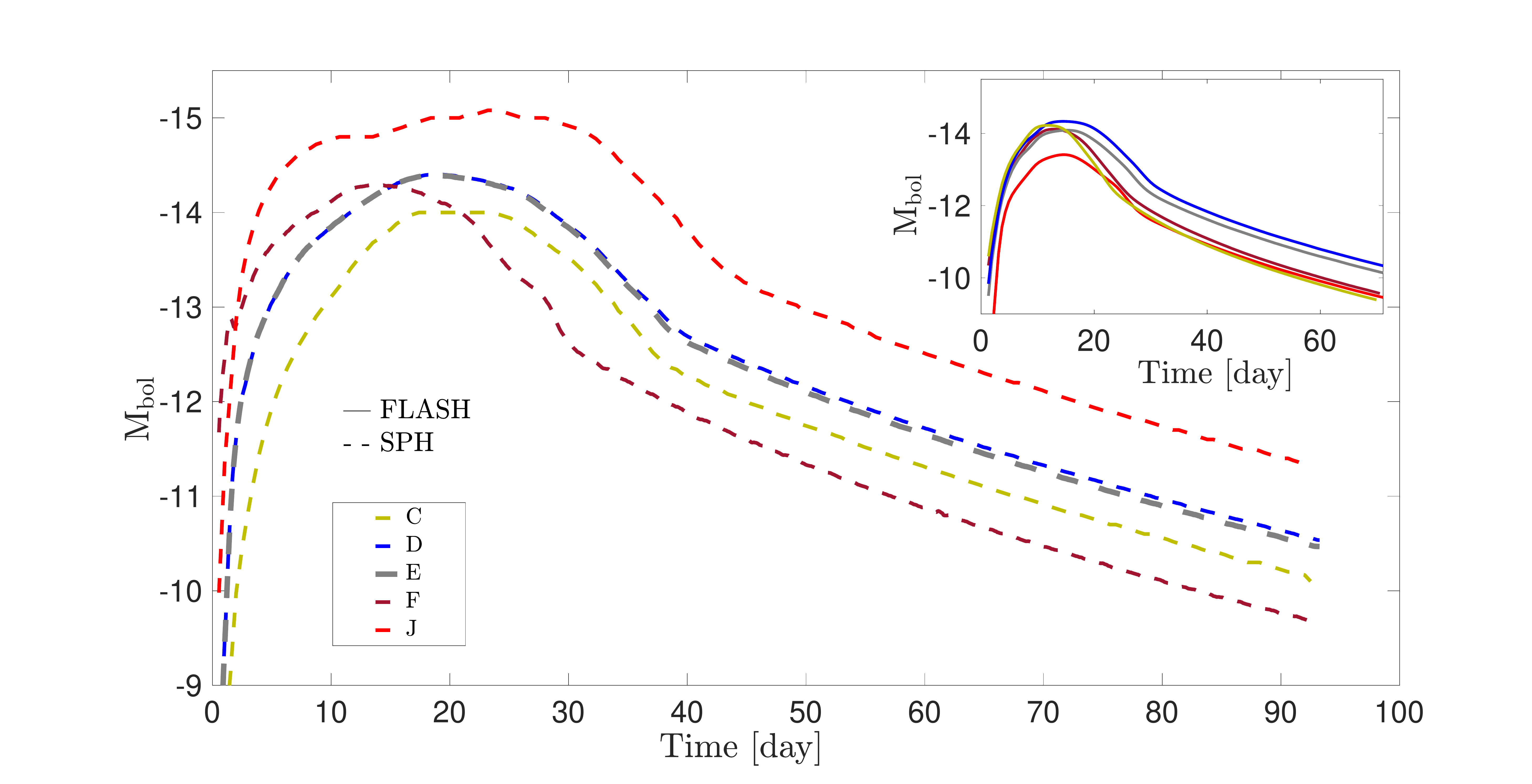}
\includegraphics[scale=0.16]{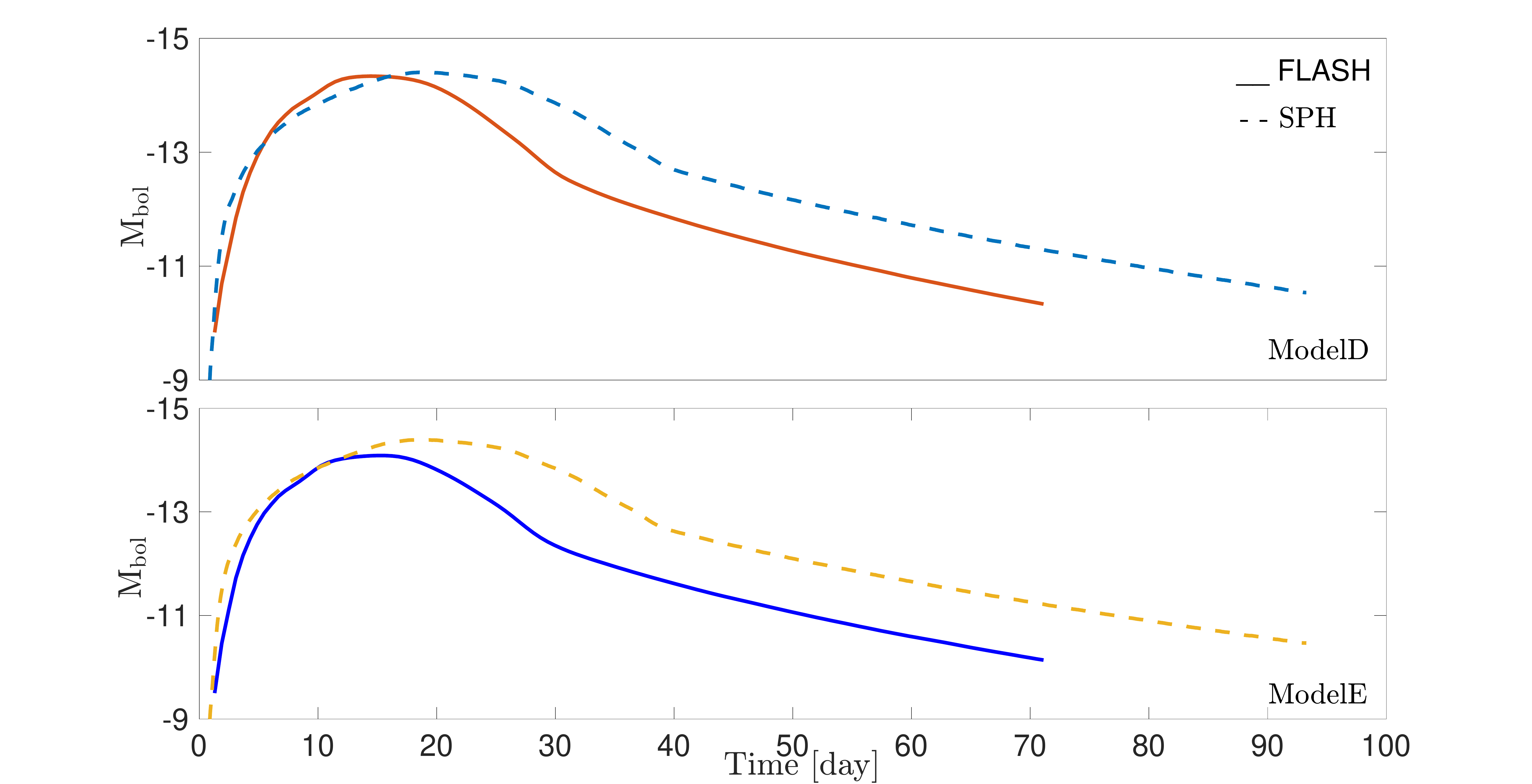}
\caption{
The bolometric lightcurves.
Top: The bolometric lightcurves for all models based on the FLASH and SPH simulations. The subplot shows the light curves for all the models based on the FLASH simulations. Bottom: Comparison of the bolometric light curves from the SPH and the FLASH models for models D and E.}
\label{fig:lightcurve0}
\end{figure}
 \begin{figure}
\includegraphics[scale=0.17]{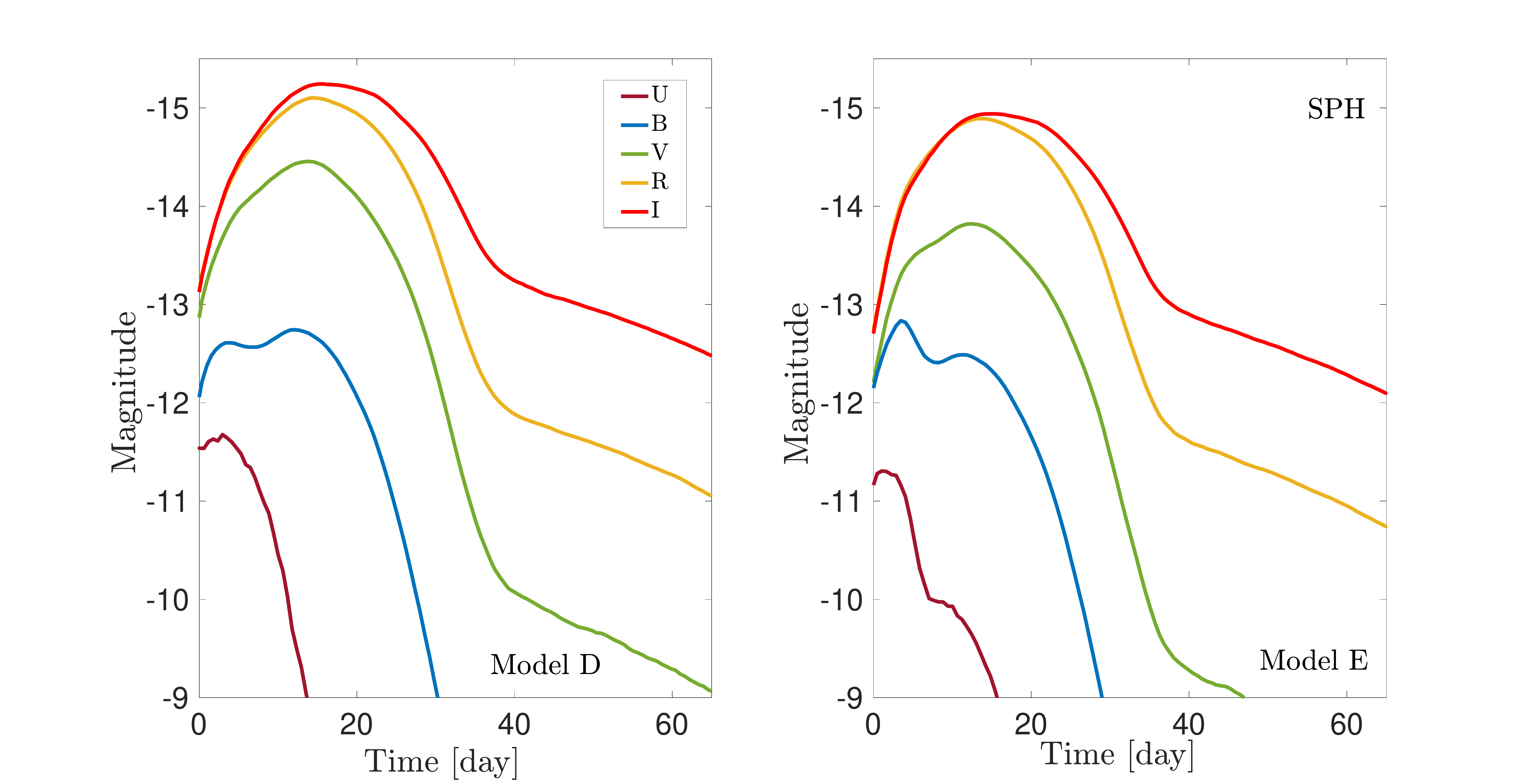}
\includegraphics[scale=0.17]{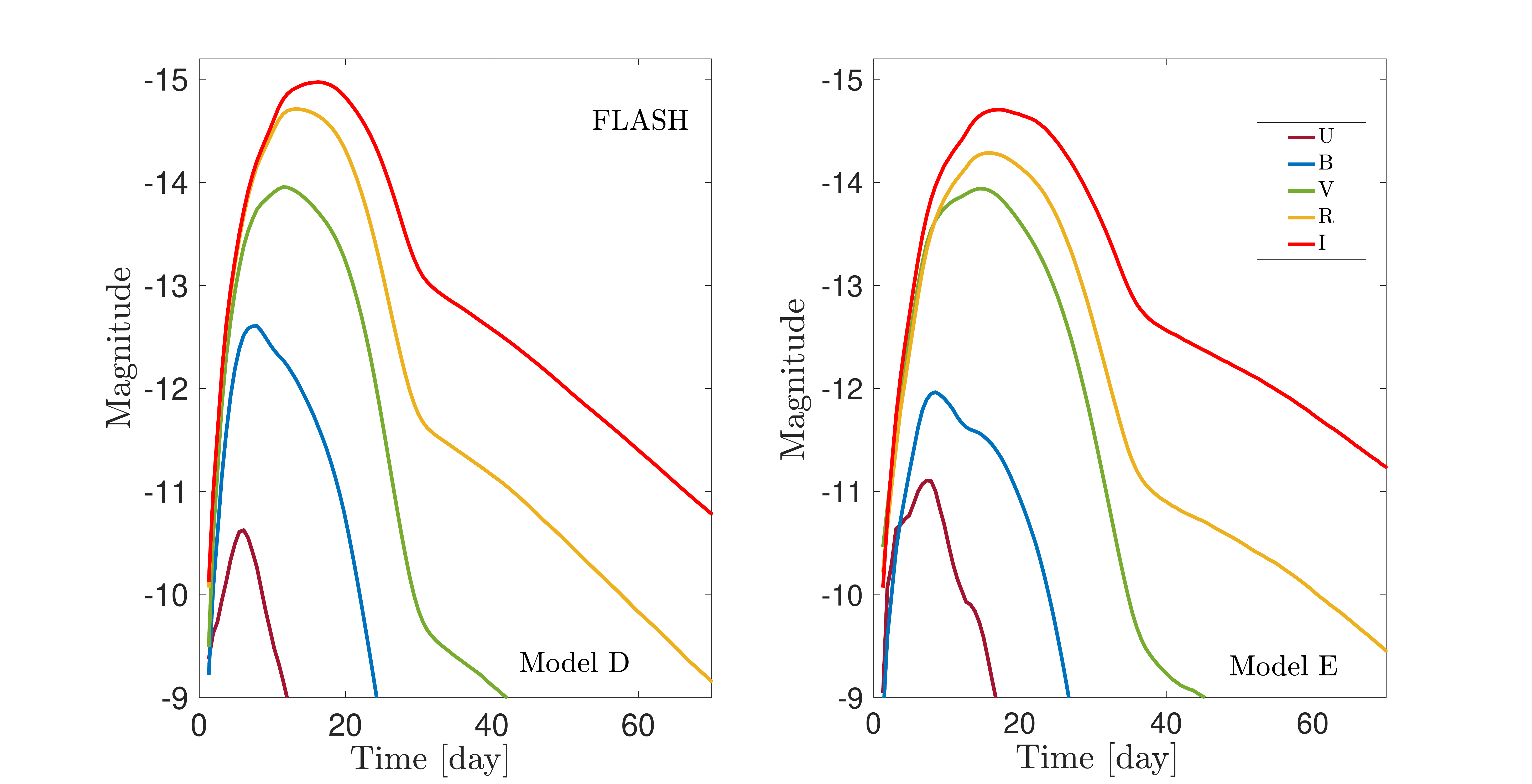}

\caption{The UBVRI light curve (LC) for model D (left) and model E (right). The top panels show the results from the radiative transfer processing of the SPH simulations. The bottom panels show for the same models from based 2D FLASH simulations.
}
\label{fig:lightcurve}
\end{figure}

\begin{figure}
\includegraphics[width=0.98\linewidth]{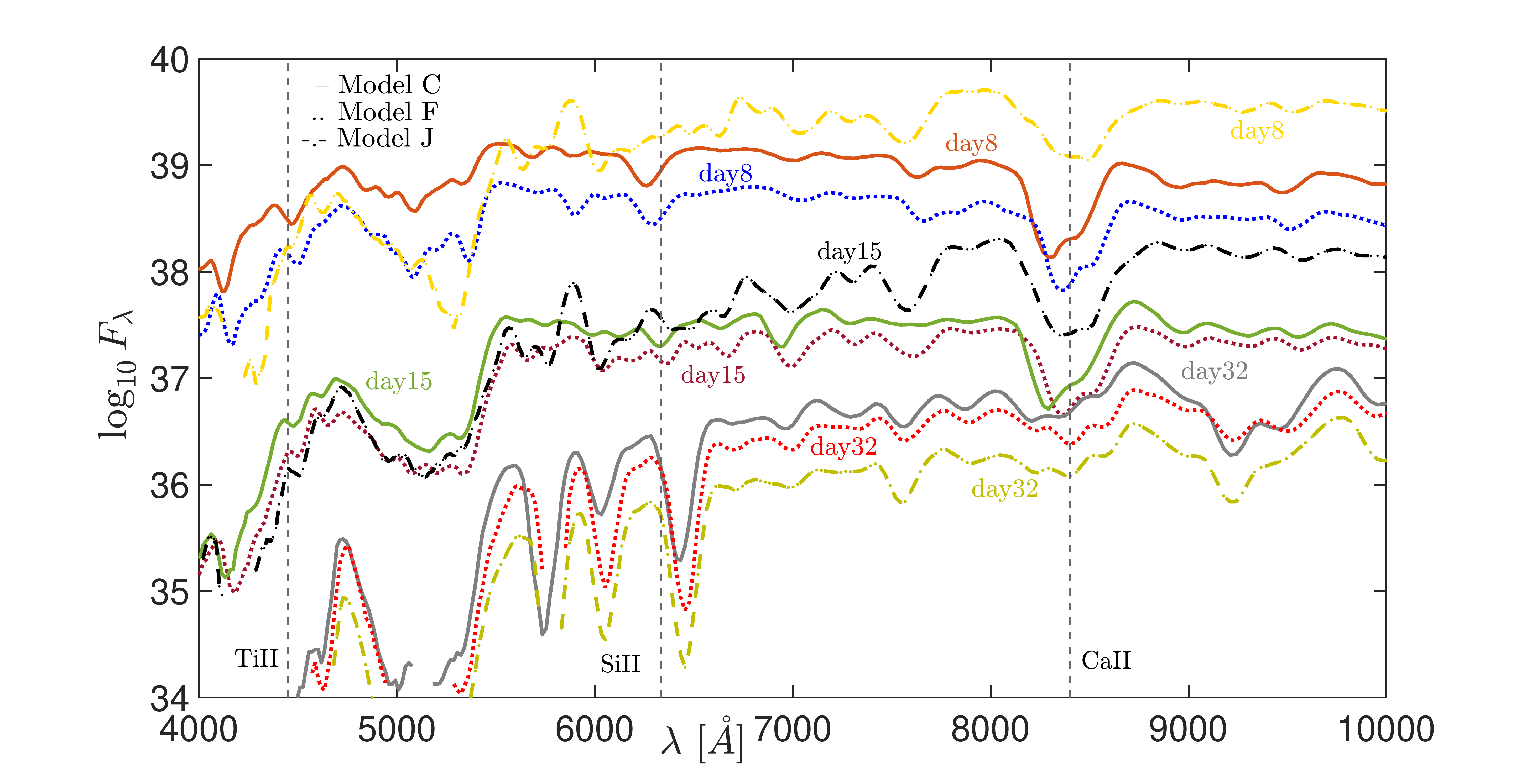}
\includegraphics[width=0.98\linewidth]{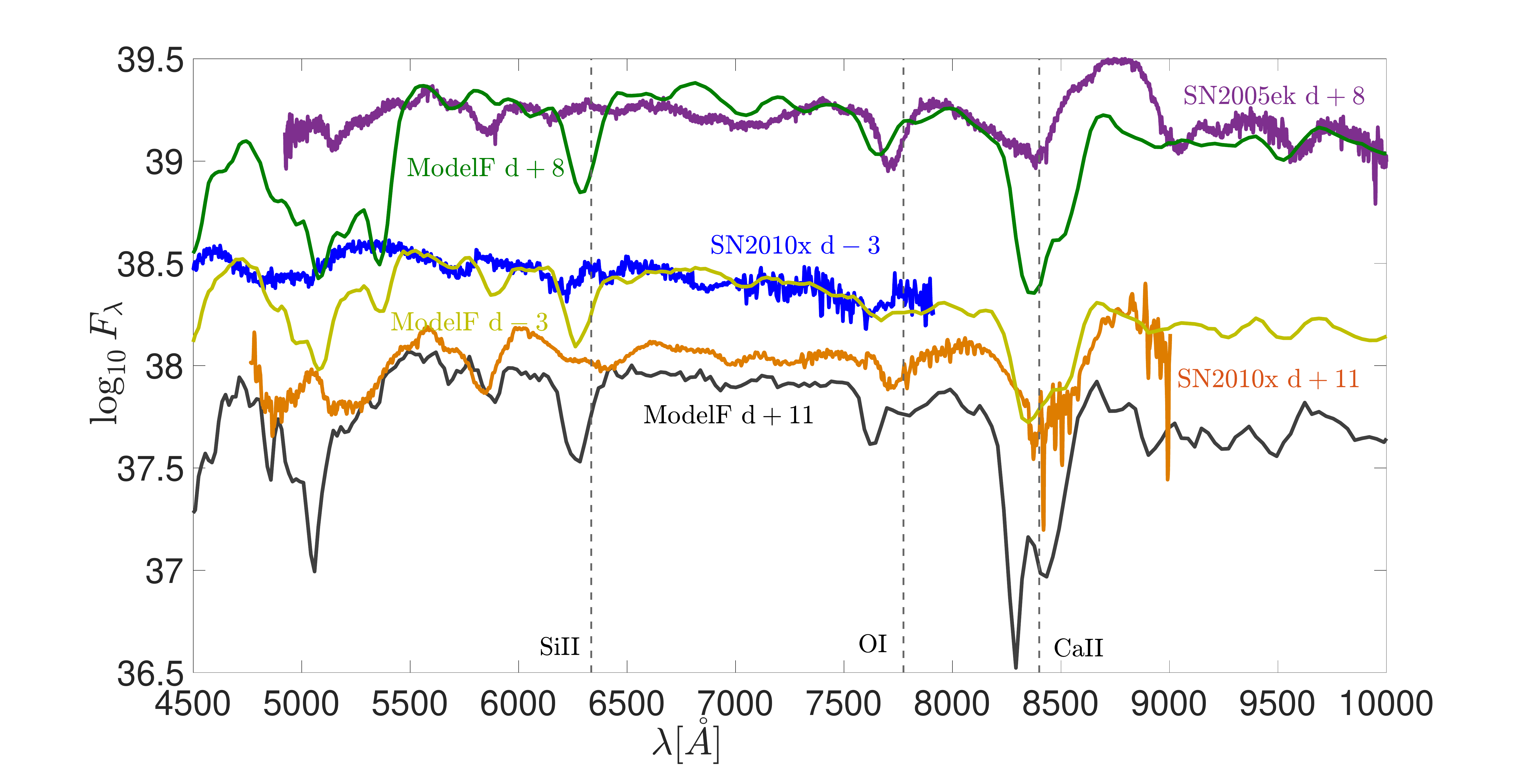}
\includegraphics[width=0.98\linewidth]{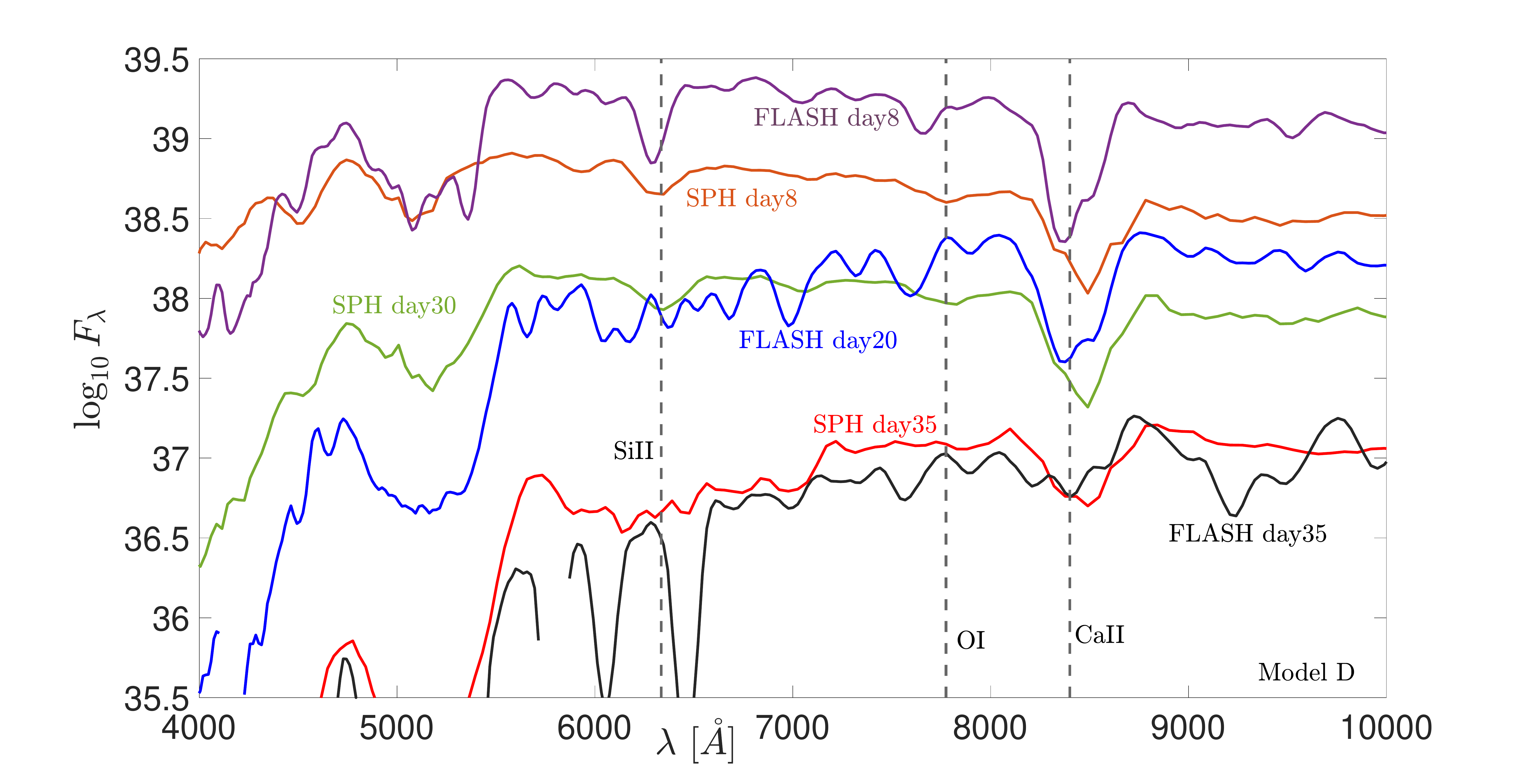}
\caption{Spectra for models C, F and J from the FLASH simulations shown at different times (top panel), spectra for model F superposed with spectra from SN2010x and SN2005ek (middle panel) and a comparison of the FLASH- and SPH-based spectra for model D (bottom panel).}
\label{fig:spectra1}
\end{figure}

\begin{figure}
\includegraphics[width=0.98\linewidth]{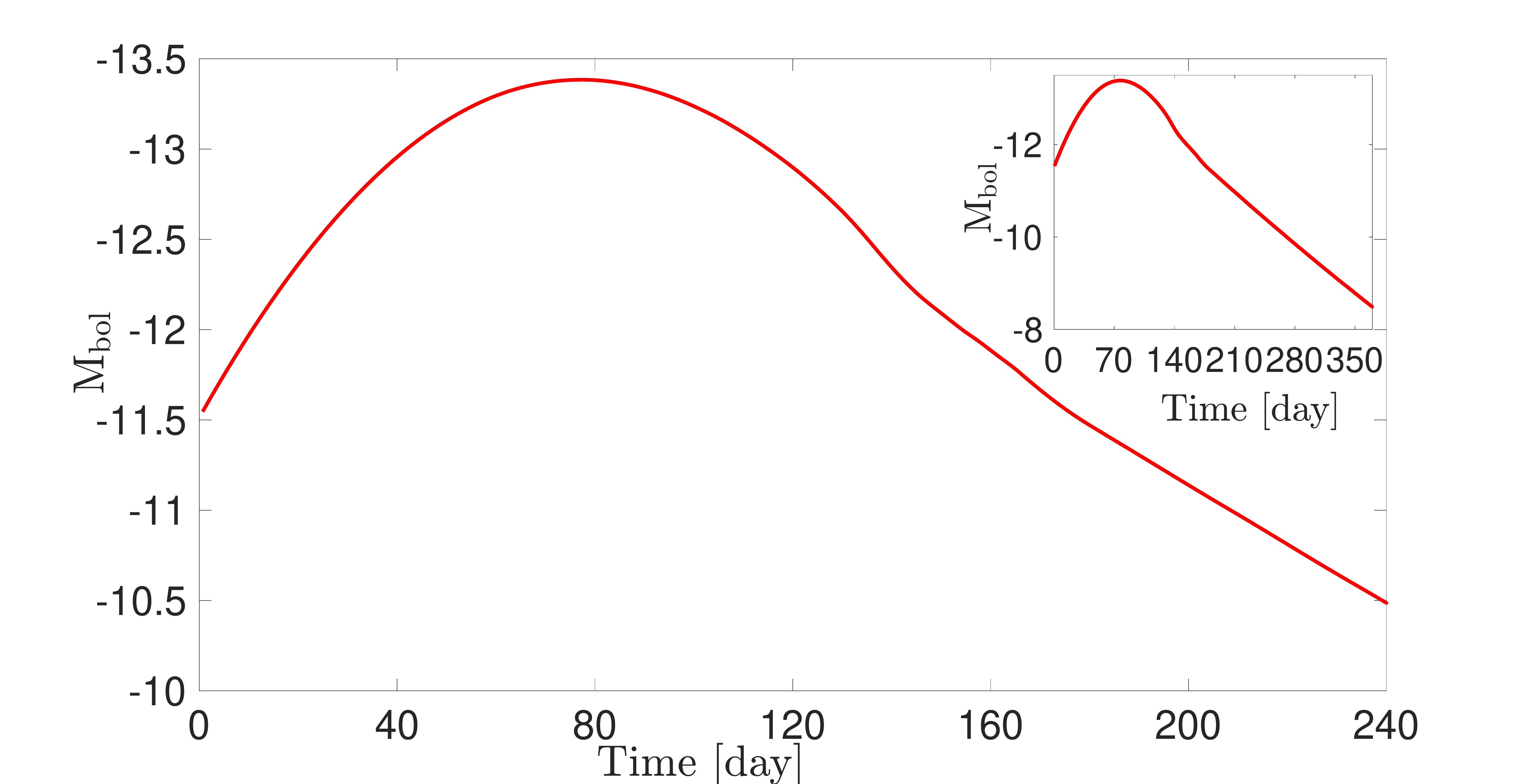}
\includegraphics[width=0.98\linewidth]{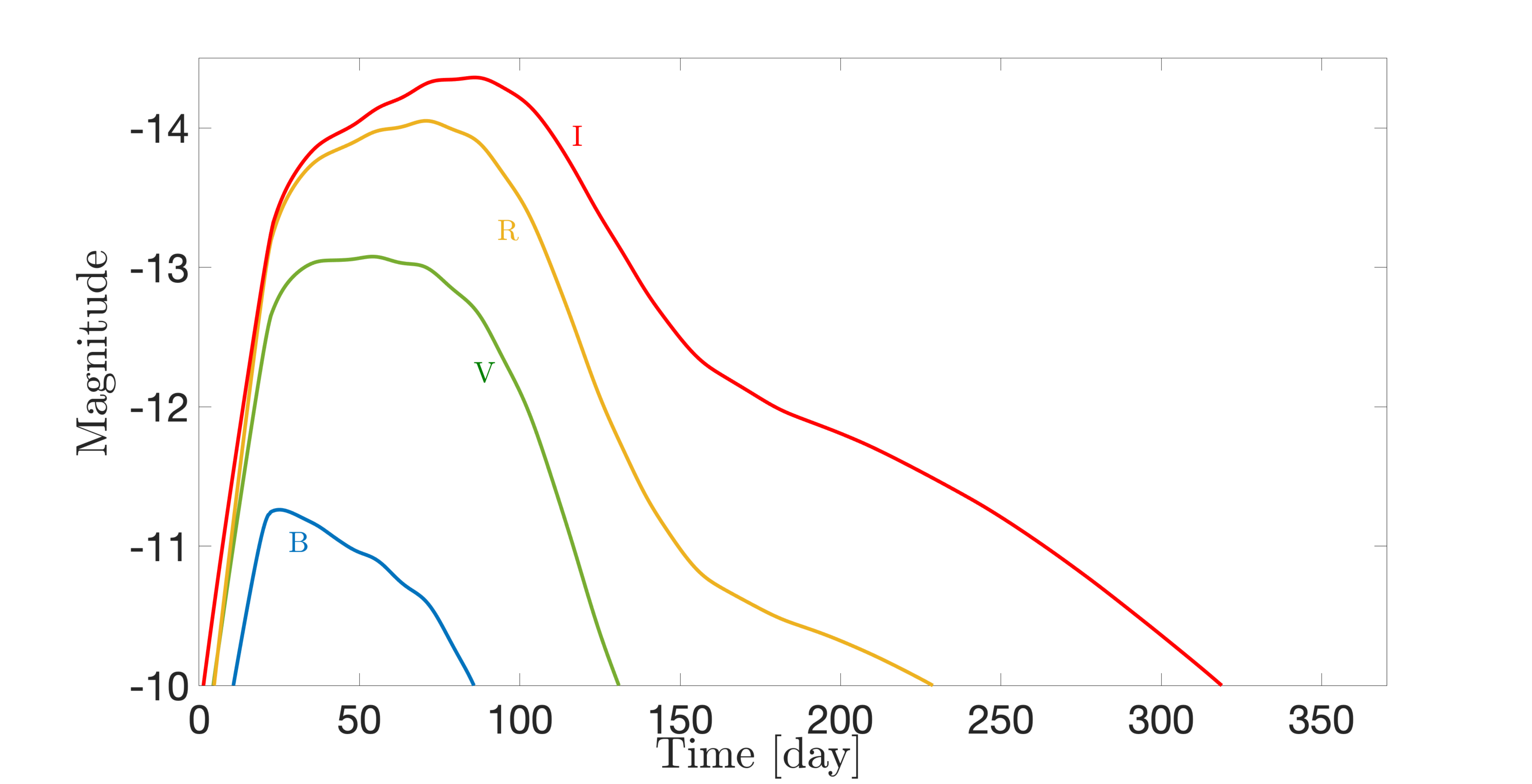}
\includegraphics[width=0.98\linewidth]{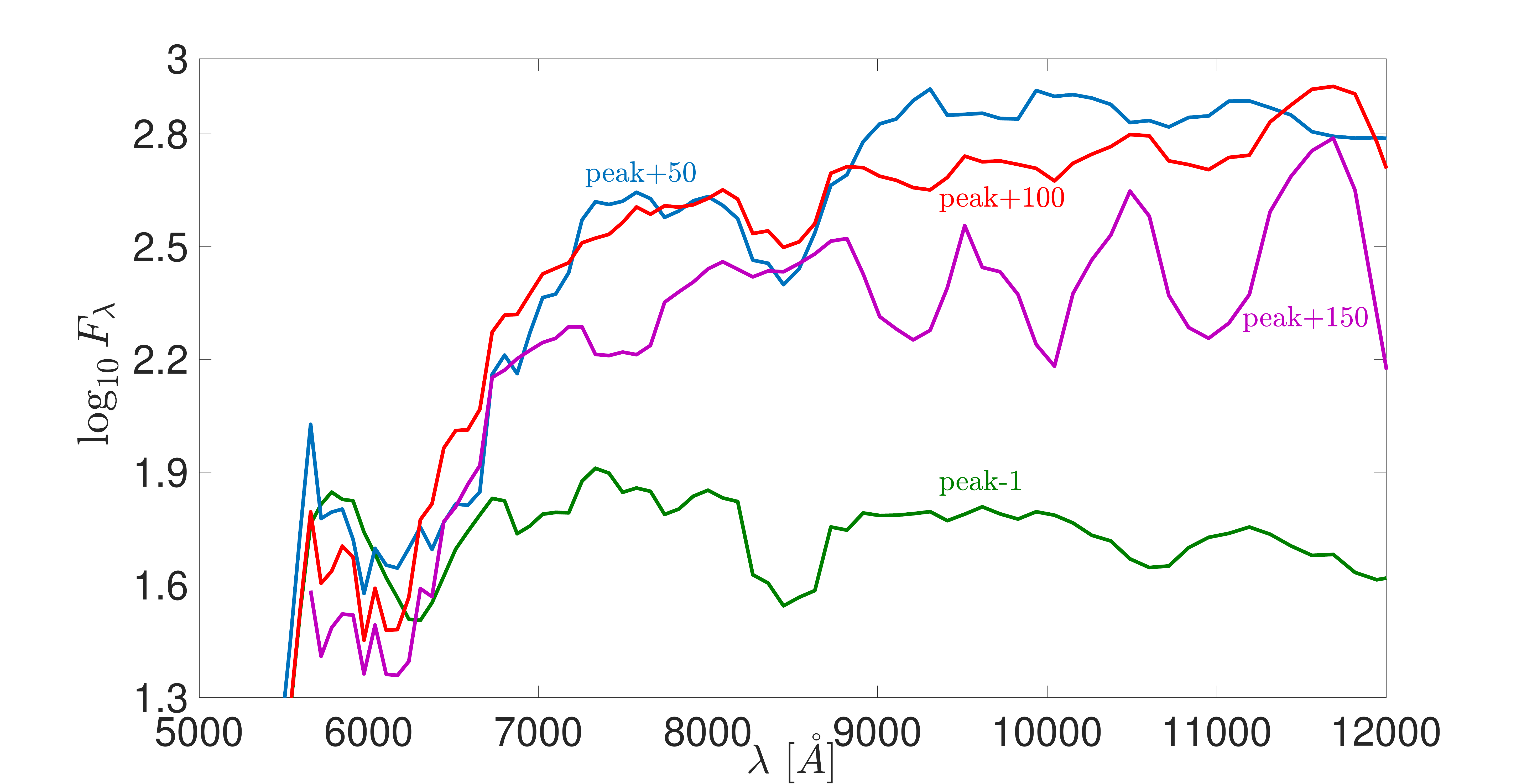}
\caption{Light curve (top panel), BVRI bands (middle panel) and spectra (bottom panel) for the largest possible secondary transient for model D, in which all the material is ejected as a slow wind $v = 400\,{\rm km}/{\rm s}$, and a $1$-$\sigma$ spread of $200\,{\rm km}/{\rm s}$. The upper panel: the light curve zoom on the first 240 days, the inset figure shows the light curve for up to 372 days. The middle panel: lightcurves in the B, V, R, I bands. U band is not shown since its luminosity is below mag $-8$. The lower panel: the spectra at $-1$, $50$, $100$, $150$ days after peak. The secondary red transient is expected to have a wide peak with half-time of $\sim{50}\,{\rm d}$, while its peak magnitude is expected to be no larger than in this example.
}
\label{fig:Afterglow}
\end{figure}

\begin{figure*}
\includegraphics[width=\linewidth]{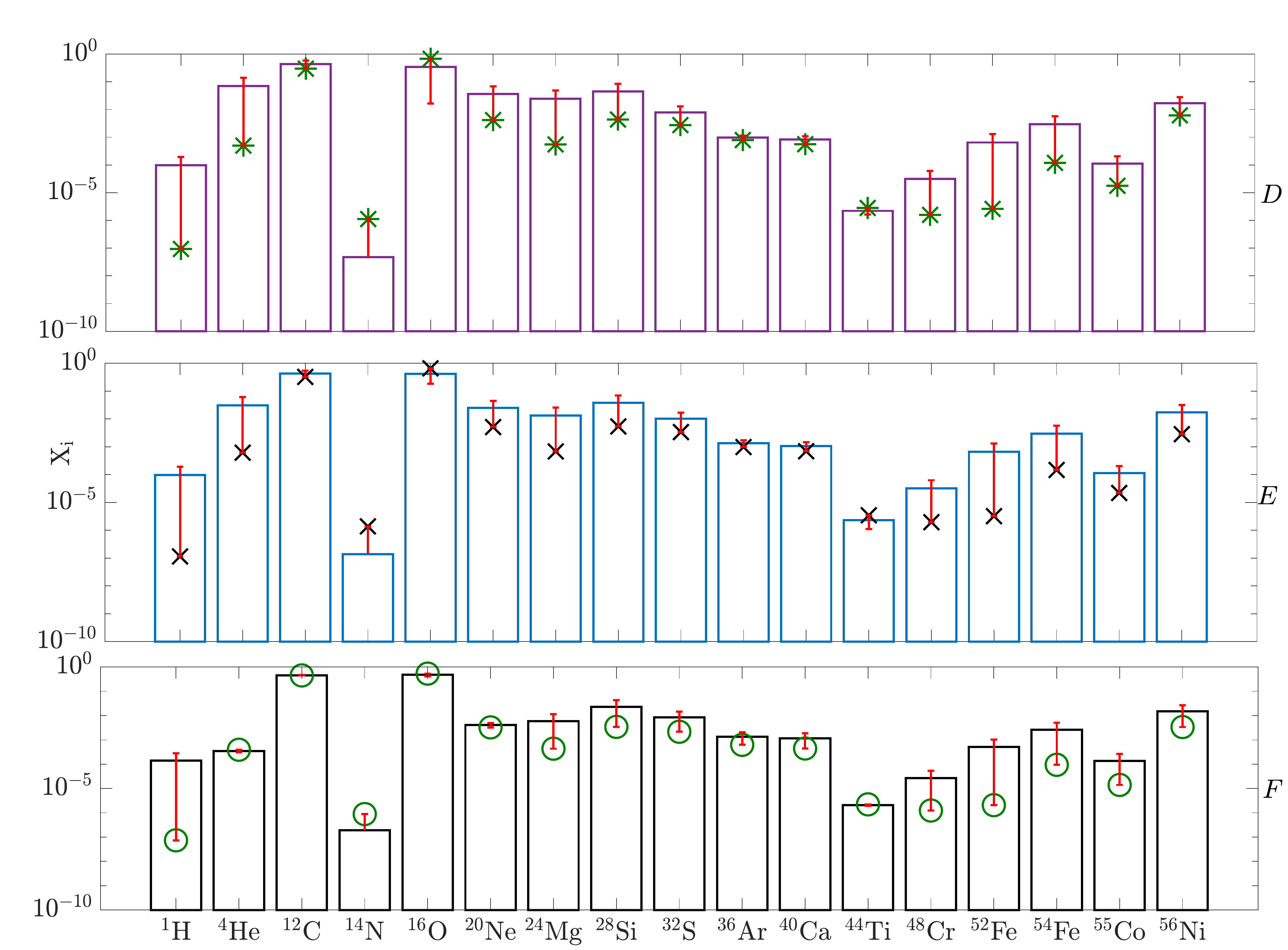}
\caption{Nuclear yields from the FLASH and the SPH simulations (star and box symbols, correspondingly), with the bars showing the difference between the SPH and the FLASH yields. The yields for 17 elements resulting from nuclear post-processing of models D, E and F show an overall good agreement: the abundances of most alpha-elements, including $^{56}{\rm Ni}$ are within $1\,{\rm dex}$ from the two codes, and the overall abundance patterns are similar.}
\label{fig:elements PPNs'}
\end{figure*}

\begin{figure*}

\includegraphics[width=\linewidth]{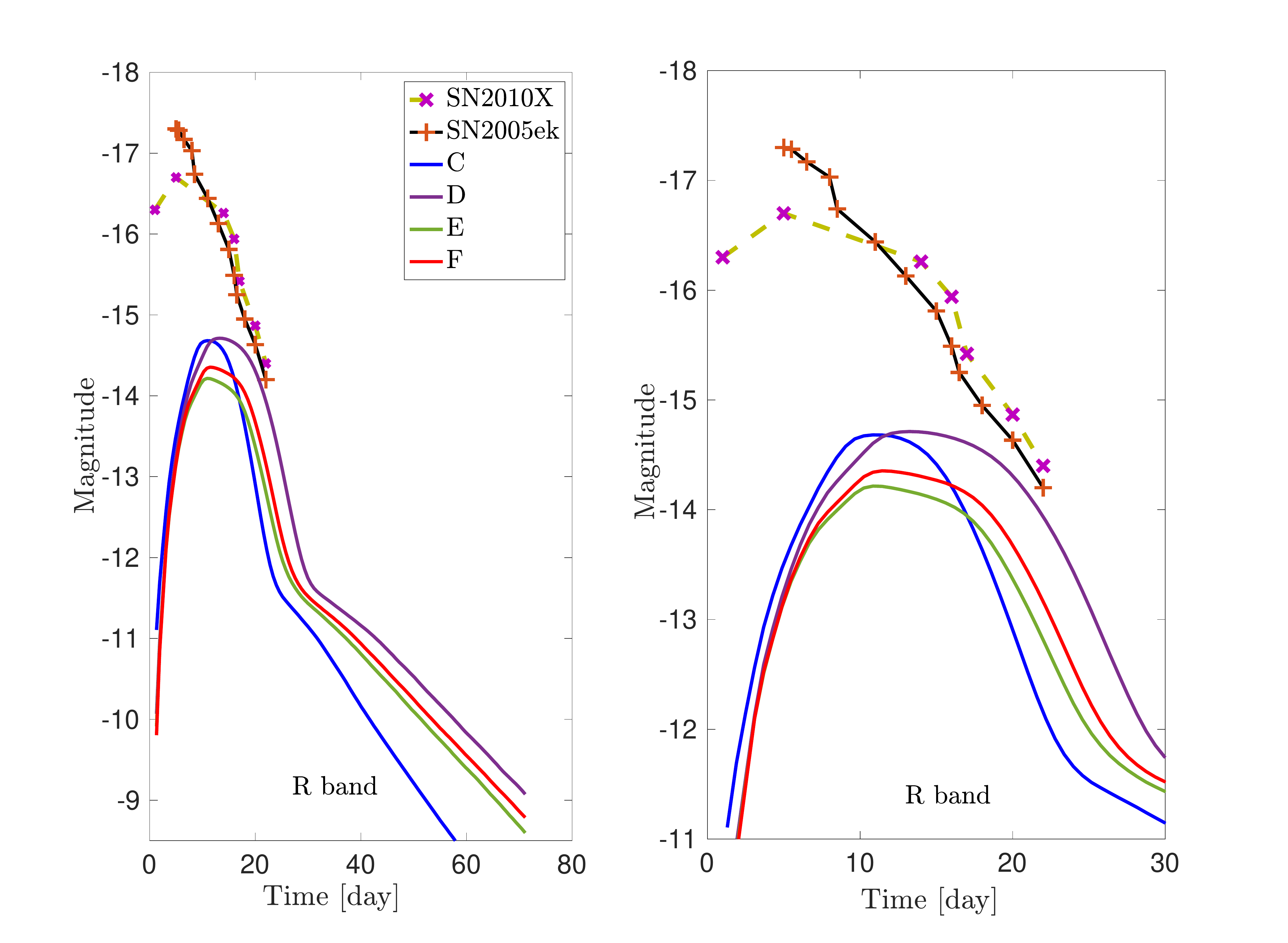} 
\caption{Comparison of the R-band lightcurves of SN2010x and SN2005ek with the modelled lightcurves based on the FLASH runs. While sharing similar durations, the rapid red transients (RRTs) resulting form NS-CO WD mergers show much fainter peak luminosities.}
\label{fig:R-band light curve'}
\end{figure*}

\section{Results}

\subsection{Merger dynamics}
\label{sec:MergDyn}

We infer the three-dimensional picture of the merger and its long-term behaviour from our SPH simulations (see Fig. \ref{fig:SPH}). Initially, the mass-transfer rates are low, and the evolution resembles that of stably transferring binaries. However, as the Roche lobe gradually cuts off increasingly large fractions of the donor, the mass transfer rate grows over time. At a certain point, the mass transfer rate is so high, that the donor effectively becomes gravitationally unbound and enters a free fall onto the accretor \cite{MM+17}. During this phase, the tidal forces stretch the white dwarf (see figure \ref{fig:SPH}), causing it to self-intersect, thus producing shocks and circularising the material into a disc. The density and temperature rise in the shocked regions, which leads to, initially mild, nuclear burning. Due to the limitations of the 2D models, our 2D FLASH models (see Fig. \ref{fig:FLASH}) begin only after the formation of the debris disk and subsequently follow its nuclear-hydrodynamic evolution.

In the FLASH models, we see that as the density and temperature increase in the inner regions, they attain critical conditions for thermonuclear burning, which leads to a weak explosive detonation.  We discuss this in more detail in Paper I. However, only a small fraction of the accreting material participates in thermonuclear burning. As a result, this produces only small amounts of nuclear by-products. Both the FLASH and the post-processing of the SPH modelling show that the mergers produce only up to  $\sim10^{-3}-10^{-2}\,{M_{\odot}}$ of intermediate or even iron group elements. In particular, the amount of $^{56}{\rm Ni}$ is always at most $10^{-2}\,{M_{\odot}}$. The bulk of the nuclear material is produced in the developed disc stages, well represented both by SPH and FLASH. The total released nuclear energy, as reported from the 19-isotope network in FLASH simulation, reaches only up to $\sim{\rm few\,\times}10^{46}-10^{47}\,{\rm erg}$. For comparison, the amount of the released gravitational energy, calculated as the total increase in the kinetic and internal energies minus the total generated nuclear energy, see, e.g. \citep{FM13,Zen+19a,FMM19}, is a few times $10^{47}$ to $10^{49}\,{\rm erg}$. Thus, the released gravitational energy is $10$ to $100$ times larger than the released nuclear energy. Hence, gravitational dynamics drives most of the outflows and heats most of the material in the systems.

Our models do not follow the long-term evolution of the NS-WD mergers and, in particular, the evolution of the fallback material that could potentially contribute to the transient luminosity. Nevertheless, we briefly discuss a simplified analytical model for the possible fallback contribution in the Appendix. We find that the fallback material is unlikely to contribute significantly to the transient energetics and the observable lightcurves.

~

~

\subsection{Light curve and spectra properties}

Figures \ref{fig:lightcurve0} and \ref{fig:lightcurve} show the resulting bolometric and UBVRI lightcurves from our models, while Fig. \ref{fig:spectra1} shows the predicted spectra from these models. Table \ref{tab:WD-models2} summarizes the basic properties of the transients.
Figure \ref{fig:Afterglow} shows the lightcurve and spectra for the possible secondary transient from the disk-wind material.

As can be seen, our models predict that NS - CO-WD mergers give rise to vary faint transients with absolute B magnitudes in the range between $-10.9$ and $-12.8$. These transients are far brighter in the I/R bands, with peak I magnitudes in the range between $-13.9$ and $-15$. These transients show extremely fast decline in the B-band ($\Delta{\rm B}_{15}$ in the range $3.35-4.5$) and rapid decline in the redder bands ($\Delta {\rm I}_{15}$ in the range $1.8-2.34$).  Such rapid red transients (RRTs), as predicted by our models, are difficult to observe in current surveys, and have not been observed yet to the best of our knowledge. Nevertheless, next-generation surveys such as the upcoming Large Synoptic Sky Survey (LSST), might be able to observe tens of such events every year (see the next section). An observational feature specific to these transients may be a faint (up to mag $-13$) $^{56}{\rm Ni}$-powered several-months long secondary red afterglow transient. Due to lower ejection velocities compared to RRTs, the expansion happens on roughly $10$ times longer timescales, resulting in larger diffusion time, later nebular phase and several month-long peak half-times. The peak luminosity of these secondary transients are an upper estimate and are likely lower, depending on the fraction of the mass lost through the disc wind ejecta.

Our results show comparable ranges of ejected masses, velocities and $^{56}{\rm Ni}$ production as found in the simplified 1D models by \citet{Mar+17}. These earlier models provided basic estimates for the bolometric light curves which are broadly in agreement with our detailed bolometric light curves.

\subsection{Rate estimates}

Despite the relatively small peak luminosities of such transients, upcoming synoptic surveys such as LSST will observe them frequently. With the $5$-$\sigma$ magnitude limits for single-visit detections of $24.0$ and $24.7$ in I and R bands, respectively \citep{LSST2009}, LSST will be able to detect such transients within $15$ days past the peak up to a distance of $250\,\textrm{Mpc}$. In order to estimate the detection rates, we assume that the merger rates of CO WD-NS binaries scale with the blue luminosity (total luminosity, e.g. of galaxies, in the B-band). By using the galactic merger rate of $1-2\cdot 10^{-4}\,\textrm{yr}^{-1}$ from our  population synthesis studies of NS-WD mergers \citep{Too+18b}, the galactic blue luminosity of $1.7\cdot 10^{10}L_{{\rm B},\odot}$ \citep{Kalogera2001} and the blue luminosity for the local universe $1.98\cdot 10^{8}\,{L_{{\rm B},\odot}}/{{\rm Mpc}^3}$ \citep{Kopparapu2008}, we find that LSST will detect between $10$ and $70$ transients per year within its field of view. The detection efficiency is similar for R and I bands. In comparison, the ongoing ZTF survey is expected to observe between $0.05$ and $0.3$ such transients per year, assuming R-band detection limit of $21$, consistent with the current null-detection of such transients by ZTF.

~

~

\subsection{Composition}
The detailed composition yields from our FLASH models have already been presented in detail in Paper I. The results from the post-processing of the SPH simulations show overall agreement with the FLASH results.  The nuclear yields for most alpha-elements, as well as $^{56}{\rm Ni}$, are within $1\,{\rm dex}$, as produced by the two codes, and the abundance patterns look similar, although suggesting systematically somewhat lesser production of burned material in FLASH. The detailed comparison for 17 elements can be seen in Fig. \ref{fig:elements PPNs'}.

\section{Discussion and summary}
\label{sec:Disc}
In our study, we explored the observable manifestation of transients from NS-WD mergers. We have made use of AMR and SPH hydrodynamical simulations to model the mergers of NSs with CO WDs, followed their thermonuclear evolution through concurrent and post-process analysis and then used radiative-transfer models to predict the expected lightcurves and spectra of such transients. We generally find that NS-WD mergers give rise to ultra-faint, rapidly evolving reddened transients (RRTs), likely observable with next-generation surveys such as the LSST.

In our study, we have qualitatively confirmed, for the first time with a 3D code, the nuclear yields in NS-WD mergers obtained in earlier axisymmetric simulations, e.g. \citep{FM13,Zen+19a,FMM19}. As discussed in Section~\ref{sec:RTSetup}, the fraction of the fast material ejected dynamically, is the major uncertainty in our model, and our synthesised lightcurves provide the maximal expected signal expected for RRT. The FLASH simulations, dominated by $\alpha$-viscosity, may be representative of the MRI-dominated disc stage following the merger and produce up to $0.1$-$0.2\,M_\odot$ of fast ejecta, thus supporting the idea that the amount of fast ejecta may be significant. The slow disc ejecta produced later on may result in a faint red several months-long afterglow. The subsequent evolution of the remnant object has been studied in detail by \citet{MM2017}.

Our predicted RRTs are much fainter and faster evolving than both typical Ia SNe as well as other peculiar faint and/or rapid transients observed to date.
Over the last two decades several classes of fast-evolving SNe had been identified  \citep{deV+85,2010Sci...327...58P,1992Tutukov,Kas+10,2011ApJ...730...89P,Dro+13,Dro+14,RuiterAshleyRe}. However, RRTs are too faint and typically redder than these SNe. Even the faintest rapidly evolving SNe differ significantly from our predicted RRTs, as can be seen in the lightcurve in figure \ref{fig:R-band light curve'} comparing SN 2010X and SN 2005ek with the RRTs. Moreover, their helium and aluminium content may be too small to explain the He and Al lines identified in these SNe \citep{Kas+10}, and they show much stronger Si lines. Other faint type Ia SNe such as Ca-rich 2005E-like SNe \citep{2010Natur.465..322P} or SN 2008ha-like SNe \citep{Foley_2009} are still much brighter in the blue bands, are not as red and decline much slower than our predicted RRTs.  

Transients from NS-WD mergers may, therefore, represent a completely different class of SNe that might be observed mostly in close-by galaxies using large telescopes or possibly with next-generation surveys such as LSST, as discussed above. As we show in \citet{Too+18b} and discuss above, the rates of NS-WD mergers could be sufficiently high as to be observed by LSST. The delay-time distribution for such mergers peaks at early times of hundreds of Myr up to one Gyr, and they are therefore expected to occur preferentially in late-type host galaxies.

Finally, as we noted in Paper I, the total masses accreted onto the NS in our models appear to be small, and we, therefore, expect that NS-WD mergers are unlikely to produce regular GRBs. Nevertheless, one cannot exclude ultra-faint long GRBs lasting over timescales comparable to accretion timescales; in this case, an ultra-long GRB accompanied by an RRT would provide a very clear smoking-gun signature for the origin of both these types of transients.

\section*{Acknowledgements}
We thank Dr. Daan Van Rossum, Prof. Silvia Toonen and Prof. Robert Fisher and Prof. Ari Laor for stimulating discussions. The simulations with the Water code and their post-processing were performed on the resources provided by the Swedish National Infrastructure for Computing (SNIC) at the Lunarc cluster.


\bibliographystyle{mnras}
\bibliography{Sne2017}
\appendix

\section*{Appendix: Late-time fallback and fallback-powered lightcurve}
A significant fraction of the mass of the disrupted WD remains bound during the entire merger. The fallback material may, therefore, accrete onto the NS and potentially contribute to the emission from the merger over long timescales after the initial disruption of the WD. However, our radiative transfer models assume a homologous expansion of the material and therefore do not account for any contribution from the fallback material.

To test how sensitive are our predictions to the way we initialize the ejecta properties in SuperNu, we, therefore, considered a simplified analytical model following \citet{Dexter_2013}. It describes how the longer-term fallback of material onto the accretor powers the evolution of the light-curve, and predicts
\begin{equation}
\label{eq:tacc}
L_{\rm fallback}\left(t\right)=L_{0}\left(\frac{t_{0}}{2t_{d}}\right)^{n}e^{-t^{2}/2t_{d}^{2}}\left(-\frac{1}{2}\right)^{n/2}\times
\end{equation}
\[
\times\left[\gamma\left(1-\frac{n}{2},-\frac{t_{0}^{2}}{2t_{d}^{2}}\right)
-\gamma\left(1-\frac{n}{2},-\frac{t^{2}}{2t_{d}^{2}}\right)\right],
\]
where $\gamma\left(s,x\right)$ is the incomplete gamma-function, $t_{0}=t_{\rm dyn}$, $n={5/3}$,  and $t_{d}$ follows from \citet{1979ApJ...230L..37A}:

\begin{equation}
\label{eq:td}
{t_{d}=\sqrt{\frac{3}{4\pi}\frac{\left(M_{\rm ejecta}+M_{\rm fallback}\right)\kappa}{v_{\rm final-ejecta}\ \cdot c}}}
\end{equation}
where $\rm c$ is speed of light; for the complete solution see \citet{Dexter_2013}. 
Using this model we find that the potential contribution to the light-curves from the fallback model are $3$~--~$7$ orders of magnitude smaller in luminosity compared to the ones predicted by our SuperNu modelling, or by our $E_{\rm SN}$ model in \citet{Zen+19a}. Given these results, we believe it is safe to neglect the contribution from the emission from the fallback accretion.    

In the longer term, the remnants form a stable thick disk-like puffy structure around the NS. In the yet longer-term, the leftover debris produce a more spherical cloudy structure around the NS. Its detailed evolution is beyond the scope of our models and is to be studied elsewhere.


\bsp	
\label{lastpage}
\end{document}